%% file: main.tex
\documentclass{nature} 

\usepackage[T1]{fontenc}
\usepackage[utf8]{inputenc}
\usepackage{textgreek}

\usepackage{helvet}    
\usepackage{sectsty}   
\allsectionsfont{\sffamily}  

\usepackage{newtxtext,newtxmath}    
\usepackage{amsmath, amssymb, bm} 
\usepackage{amsthm}               
\usepackage{graphicx}             
\usepackage{xspace}               
\usepackage{booktabs}             
\usepackage{multirow}
\usepackage{pdflscape}            
\usepackage[left=1in,right=1in,bottom=1.1in,top=1in]{geometry}
\usepackage{enumitem}             
\usepackage[table]{xcolor}        
\usepackage[dvipsnames]{xcolor}   
\usepackage{siunitx}              
\usepackage{textcomp}
\usepackage[normalem]{ulem}       
\usepackage[ruled,vlined,linesnumbered]{algorithm2e} 
\usepackage[innercaption]{sidecap} 
\usepackage{lineno}               
\usepackage{verbatim}             
\let\spacing\relax
\usepackage{setspace}             
\usepackage{xr}                   
\usepackage[colorlinks,citecolor=blue,urlcolor=blue]{hyperref} 
\usepackage{xr-hyper}
\usepackage{float}
\usepackage{url}
\usepackage{orcidlink}

\DeclareSIUnit\Molar{M}
\DeclareSIUnit\rpm{rpm}
\DeclareSIUnit\ppm{ppm}

\usepackage[natbib=true, citestyle=numeric-comp,
            bibstyle=nature, sorting=none,
            maxbibnames=6, doi=true, url=false, isbn=false, date=year]{biblatex}

\AtEveryBibitem{%
  \clearfield{note}%
  \clearlist{note}%
  \clearfield{language}%
  \clearlist{language}%
}

\usepackage[font=footnotesize, labelfont={sf,bf}]{caption}
\DeclareCaptionLabelFormat{extended_data}{\textbf{Extended Data #1 #2: }}

\graphicspath{{FIG/}}

\newcommand{\xhdr}[1]{\vspace{1.2mm}\noindent\textsf{\textbf{{#1.}}}\xspace}

\newcommand{\eg}{\emph{e.g.}\xspace}
\newcommand{\ie}{\emph{i.e.}\xspace}
\newcommand\etal{\emph{et al.}\xspace}

\newcommand{\textsbf}[1]{\textsf{\textbf{#1}}}

\SetKwRepeat{Do}{do}{while}
\let\oldnl\nl
\newcommand{\nonl}{\renewcommand{\nl}{\let\nl\oldnl}}

\theoremstyle{definition}

\newcommand\model{\textsc{Proton}\xspace}
\newcommand\kg{\textsc{NeuroKG}\xspace}

\newcommand{\scriptG}{\mathcal{G}}
\newcommand{\scriptV}{\mathcal{V}}
\newcommand{\scriptE}{\mathcal{E}}
\newcommand{\scriptA}{\mathcal{A}}
\newcommand{\scriptR}{\mathcal{R}}
\newcommand{\scriptN}{\mathcal{N}}
\newcommand{\scriptD}{\mathcal{D}}

\title{
\begin{center}
Graph AI generates neurological hypotheses validated \\in molecular, organoid, and clinical systems
\vspace{-10mm}
\end{center}
}

\author{
\begin{center}
Ayush~Noori$^{1,2,3,4,5,6,7}$~\orcidlink{0000-0003-1420-1236},
Joaqu\'{\i}n~Polonuer$^{1}$~\orcidlink{0009-0007-8613-6126},
Katharina~Meyer$^{2,5,8}$~\orcidlink{0000-0001-9051-1354},
Bogdan~Budnik$^{2,5}$~\orcidlink{0000-0003-3622-2003},
Shad~Morton$^{2,5}$~\orcidlink{0009-0003-1233-3785},
Xinyuan~Wang$^{6,9}$~\orcidlink{0000-0002-3107-8359},
Sumaiya~Nazeen$^{6,9}$~\orcidlink{0000-0002-6313-6357},
Yingnan~He$^{3}$~\orcidlink{0009-0003-6082-3893},
I\~naki~Arango$^{1}$~\orcidlink{0009-0002-1443-2325},
Lucas~Vittor$^{1}$~\orcidlink{0009-0002-6978-0482},
Matthew~Woodworth$^{2,5,8}$~\orcidlink{0009-0008-8255-6817},
Richard~C.~Krolewski$^{6,9}$~\orcidlink{0000-0001-5974-079X},
Michelle~M.~Li$^{1,6}$~\orcidlink{0000-0003-0223-7485},
Ninning~Liu$^{2,5}$~\orcidlink{0000-0002-8398-9584},
Tushar~Kamath$^{10}$,
Evan~Macosko$^{10}$~\orcidlink{0000-0002-2794-5165},
Dylan~Ritter$^{6,11}$~\orcidlink{0000-0002-8704-908X},
Jalwa~Afroz$^{6,11}$~\orcidlink{0009-0007-5186-6401},
Alexander~B.~H.~Henderson$^{3,6}$,
Lorenz~Studer$^{6,11}$~\orcidlink{0000-0003-0741-7987},
Samuel~G.~Rodriques$^{12}$~\orcidlink{0000-0002-2509-0861},
Andrew~White$^{12}$~\orcidlink{0000-0002-6647-3965},
Noa~Dagan$^{7,13,14}$~\orcidlink{0000-0001-8811-7825},
David~A.~Clifton$^{4,15}$~\orcidlink{0000-0002-9848-8555},
George~M.~Church$^{2,5,8}$~\orcidlink{0000-0001-6232-9969},
Sudeshna~Das$^{3,\dagger}$~\orcidlink{0000-0002-9486-6811},
Jenny~M.~Tam$^{2,5,8,\dagger}$~\orcidlink{0000-0002-3767-7205},
Vikram~Khurana$^{6,9,10,16,\dagger}$~\orcidlink{0000-0002-4018-5527},
Marinka~Zitnik$^{1,6,7,10,17,18,\dagger}$~\orcidlink{0000-0001-8530-7228} \\[4mm]
\footnotesize{$^{1}$Department of Biomedical Informatics, Harvard Medical School, Boston, MA, USA} \\
\footnotesize{$^{2}$Wyss Institute for Biologically Inspired Engineering at Harvard University, Boston, MA, USA} \\
\footnotesize{$^{3}$Department of Neurology, Massachusetts General Hospital, Boston, MA, USA} \\
\footnotesize{$^{4}$Department of Engineering Science, University of Oxford, Oxford, UK} \\
\footnotesize{$^{5}$BD\textsuperscript{2}: Breakthrough Discoveries for thriving with Bipolar Disorder, Santa Monica, CA, USA} \\
\footnotesize{$^{6}$Aligning Science Across Parkinson's (ASAP) Collaborative Research Network, Chevy Chase, MD, USA} \\
\footnotesize{$^{7}$The Ivan and Francesca Berkowitz Family Living Laboratory Collaboration at \\ Harvard Medical School and Clalit Research Institute, Boston, MA, USA} \\
\footnotesize{$^{8}$Department of Genetics, Harvard Medical School, Boston, MA, USA} \\
\footnotesize{$^{9}$Department of Neurology, Brigham and Women’s Hospital, Boston, MA, USA} \\
\footnotesize{$^{10}$Broad Institute of MIT and Harvard, Cambridge, MA, USA} \\
\footnotesize{$^{11}$The Center for Stem Cell Biology, Memorial Sloan Kettering Cancer Center, New York, NY, USA} \\
\footnotesize{$^{12}$FutureHouse Inc., San Francisco, CA, USA} \\
\footnotesize{$^{13}$Clalit Research Institute, Innovation Division, Clalit Health Services, Ramat-Gan, Israel} \\
\footnotesize{$^{14}$Faculty of  Computer and Information Science, Ben Gurion University of the Negev, Be'er Sheva, Israel} \\
\footnotesize{$^{15}$}Oxford Suzhou Centre for Advanced Research, University of Oxford, Suzhou, Jiangsu, China\\
\footnotesize{$^{16}$Harvard Stem Cell Institute, Cambridge, MA, USA} \\
\footnotesize{$^{17}$Kempner Institute for the Study of Natural and Artificial Intelligence, Harvard University, MA, USA \!\!\!\!} \\
\footnotesize{$^{18}$Harvard Data Science Initiative, Cambridge, MA, USA} \\[2mm]
\footnotesize{$^{\dagger}$Correspondence: \href{mailto:sdas5@mgh.harvard.edu}{sdas5@mgh.harvard.edu}, \href{mailto:jenny.tam@wyss.harvard.edu}{jenny.tam@wyss.harvard.edu}, \\ \href{mailto:vkhurana@bwh.harvard.edu}{vkhurana@bwh.harvard.edu}, \href{mailto:marinka@hms.harvard.edu}{marinka@hms.harvard.edu} (lead contact)} \\[2mm]
\footnotesize{\model website: \url{https://protonmodel.ai} \\
\model code: \url{http://github.com/mims-harvard/PROTON} \\
\model model: \url{https://huggingface.co/mims-harvard/PROTON}}
\end{center}
}

\begin{document}
\maketitle

\vspace{1em}
\begin{spacing}{1}
\small
\begin{abstract}
\section*{Abstract}
\input{000abstract}
\end{abstract}
\end{spacing}

\begin{spacing}{1.3}

\section*{Main}
\input{010intro}

\section*{Results}
\input{020results}

\section*{Discussion}
\input{030discussion}
\clearpage
\end{spacing}


\begin{spacing}{1}
\xhdr{Acknowledgements}
We thank Nada Amin, Uri Manor, Gabriel Kreiman, Lei Clifton, Payal Chandak, Nic Fishman, and Praveen Patnaik for helpful discussions on parts of this manuscript. We are grateful to Michael Skarlinski, Michaela Hinks, and Samantha Cox for support with the PaperQA2 API. We also thank members of the Deep Graph Library (DGL) team in the Amazon Web Services AI Shanghai Lablet (ASAIL), including Hongzhi (Steve) Chen, Minjie Wang, and Mufei Li, for support with DGL development. Finally, we thank Matias González Fernández for his support in developing the project website.

We gratefully acknowledge the support of NIH R01-HD108794, NSF CAREER 2339524, U.S. DoD FA8702-15-D-0001, ARPA-H Biomedical Data Fabric (BDF) Toolbox Program, Harvard Data Science Initiative, Amazon Faculty Research, Google Research Scholar Program, AstraZeneca Research, Roche Alliance with Distinguished Scientists (ROADS) Program, Sanofi iDEA-iTECH Award, GlaxoSmithKline Award, Boehringer Ingelheim Award, Merck Award, Optum AI Research Collaboration Award, Pfizer Research, Gates Foundation (INV-079038), Aligning Science Across Parkinson's Initiative (ASAP), Chan Zuckerberg Initiative, John and Virginia Kaneb Fellowship at Harvard Medical School, Biswas Computational Biology Initiative in partnership with the Milken Institute, Harvard Medical School Dean’s Innovation Fund for the Use of Artificial Intelligence, and the Kempner Institute for the Study of Natural and Artificial Intelligence at Harvard University. 
A.N. was supported by the Rhodes Scholarship, Dr. Susanne E. Churchill Summer Institute in Biomedical Informatics at Harvard Medical School, TIME Initiative Fellowship, and Y Combinator Summer Fellows Grant, as well as the following awards at Harvard College: the Banga Social Innovation Thesis Award, Yun Family Research Fellows Fund for Revolutionary Thinking, Herchel Smith-Harvard Undergraduate Science Research Program, and Harvard College Research Program.
D.A.C. was funded by an NIHR Research Professorship, Royal Academy of Engineering Research Chair, and the InnoHK Hong Kong Centre for Cerebro-Cardiovascular Engineering, and was supported by the National Institute for Health Research Oxford Biomedical Research Centre and the Pandemic Sciences Institute at the University of Oxford. 
V.K., S.N., and X.W. were supported by the Aligning Science Across Parkinson's Initiative (ASAP-000472). The project received generous support from the Ocko Family (S.N. and V.K.). X.W. gratefully acknowledges support from NIH T32-AG000222.
S.D. and Y.H. were supported by NIH R01-AG082698.
A.N., K.M., B.B., S.M., M.W., N.L., G.C., and J.T. gratefully acknowledge the support of BD\textsuperscript{2}: Breakthrough Discoveries for thriving with Bipolar Disorder (DG230420 and CG240420). This research was funded in whole or in part by Aligning Science Across Parkinson's through the Michael J. Fox Foundation for Parkinson’s Research (MJFF). For the purpose of open access, the author has applied a CC BY 4.0 public copyright license to this manuscript.

This research was enabled by the AI Cluster at the Kempner Institute for the Study of Natural and Artificial Intelligence at Harvard University, as well as the O2 high-performance compute cluster, supported by the Research Computing Group at Harvard Medical School. Compute was also provided by FutureHouse via the PaperQA2 API. Figures \ref{fig:overview}, \ref{fig:pd}, \ref{fig:bd}, and \ref{fig:ad}, as well as Supplementary Figure \ref{si:fig:siletti}, were created in part with Biorender.com. Any opinions, findings, conclusions, or recommendations expressed in this material are those of the authors and do not necessarily reflect the views of the funders.
\clearpage

\xhdr{Data availability} The \model project website is available at \url{https://protonmodel.ai}. \kg is available via Harvard Dataverse at \url{https://doi.org/10.7910/DVN/ZDLS3K}. Mass spectrometry data from patient-derived brain organoids are available via the MassIVE repository at \url{https://doi.org/10.25345/C5C24R16K}. This study was approved by the MGB Institutional Review Board (IRB Protocol 2015P001915).

\xhdr{Code availability}
All code used in this study is made publicly available at \url{https://github.com/mims-harvard/PROTON} under an open-source MIT license. Code to analyze the 20 original datasets from Chandak \textit{et al.}~\autocite{chandak_building_2023} is available at \url{https://github.com/mims-harvard/PrimeKG}. WASP code to analyze proteomics data is available at \url{https://github.com/Wyss-BD2/WASP3}. Model weights are released at \url{https://huggingface.co/mims-harvard/PROTON}. Analyses were conducted with Python version 3.11.10 and R version 4.4.0.

\xhdr{Supplementary information} Supplementary information is available at \url{https://github.com/mims-harvard/PROTON/tree/main/data/paper}.

\xhdr{Author contributions}
A.N. and M.Z. conceived the study.
A.N. and M.Z. designed the experiments, together with V.K. for PD-related experiments; K.M., B.B., J.M.T., and G.M.C. for BD-related experiments; and S.D. for AD-related experiments.
A.N. collected and processed the \kg pre-training data.
A.N. developed the \model model as well as the training, inference, and evaluation codebase, with support from J.P., I.A., L.V., and M.M.L.
A.N. conducted the experiments and interpreted the results, together with X.W., S.N., and R.C.K for PD-related experiments; K.M., S.M., and M.W. for BD-related experiments; and Y.H. for AD-related experiments.
D.R., J.A., A.B.H.H., and L.S. conducted the genome-wide CRISPR/Cas9 screen in midbrain dopaminergic neurons and analyzed the results.
N.L., M.M.L, N.D., G.M.C., and D.A.C. provided feedback on the analyses. T.K. and E.M. provided and analyzed snRNA-seq data from patients with PD and matched controls. S.G.R. and A.W. supported PaperQA2 analyses.
A.N., K.M., B.B., S.D., and M.Z. wrote the manuscript with input from all co-authors. D.A.C., G.M.C., S.D., J.M.T., V.K., and M.Z. supervised the research, with M.Z. leading the study.

\xhdr{Competing interests}
R.C.K. serves as an expert consultant and expert witness in litigation related to environmental toxicant exposure and Parkinson's disease, and is a topic editor in \textit{Movement Disorders} for EBSCO DynaMed. G.M.C. has a financial interest in Lila.ai and Glottatech.com. V.K. is a co-founder of and senior advisor to DaCapo Brainscience, a company focused on CNS drug discovery. 
\end{spacing}

\clearpage

\begin{spacing}{1}
\small
\input{200figures} 
\end{spacing}

\clearpage

\section*{Online Methods}
\begin{spacing}{1.3}
\input{100methods}
\end{spacing}

\clearpage

\section*{References}
\vspace{1em}
\begin{spacing}{1}
\printbibliography[heading=none]
\end{spacing}

\end{document}

%% file: 000abstract.tex
Neurological diseases are the leading global cause of disability, yet most lack disease-modifying treatments. We present \model, a heterogeneous graph transformer that generates testable hypotheses across molecular, organoid, and clinical systems. To evaluate \model, we apply it to Parkinson's disease (PD), bipolar disorder (BD), and Alzheimer's disease (AD). In PD, \model linked genetic risk loci to genes essential for dopaminergic neuron survival and predicted pesticides toxic to patient-derived neurons, including the insecticide endosulfan, which ranked within the top 1.29\% of predictions. \textit{In silico} screens performed by \model reproduced six genome-wide $\bm{\alpha}$-synuclein experiments, including a split-ubiquitin yeast two-hybrid system (normalized enrichment score [NES] = 2.30, FDR-adjusted $\bm{p < 1 \times 10^{-4}}$), an ascorbate peroxidase proximity labeling assay (NES = 2.16, FDR $\bm{< 1 \times 10^{-4}}$), and a high-depth targeted exome sequencing study in 496 synucleinopathy patients (NES = 2.13, FDR $\bm{< 1 \times 10^{-4}}$). In BD, \model predicted calcitriol as a candidate drug that reversed proteomic alterations observed in cortical organoids derived from BD patients. In AD, we evaluated \model predictions in health records from $\bm{n = 610{,}524}$ patients at Mass General Brigham, confirming that five \model-predicted drugs were associated with reduced seven-year dementia risk (minimum hazard ratio = 0.63, 95\% CI: 0.53–0.75, $\bm{p < 1 \times 10^{-7}}$). \model generated neurological hypotheses that were evaluated across molecular, organoid, and clinical systems, defining a path for AI-driven discovery in neurological disease.

%% file: 010intro.tex
Neurological diseases affect more than 3 billion people worldwide, cause over 11 million deaths each year, account for roughly one in three years lived with disability, and contribute to more than 400 million prevalent cases and tens of millions of new cases annually~\autocite{steinmetz_global_2024, feigin2020global}. Yet for most neurological diseases, there are no treatments that slow or halt disease progression~\autocite{feustel_risks_2020}. AI holds promise to accelerate the development of new therapies~\autocite{zitnik2025ai}, but AI predictions are rarely tested in disease-relevant systems that quantify their therapeutic potential. AI models are evaluated on computational benchmarks~\autocite{wang2023scientific,bunne2024build,guo2025deepseek,mcduff2025towards,shmatko2025learning}, an approach that measures internal model performance but does not test at scale whether AI predictions can translate into biological insight~\autocite{schaffer2025multimodal}. A challenge is how to develop disease-modifying treatments for neurological disease by effectively coupling AI with molecular, cellular, and clinical systems so that AI predictions can be prospectively generated and experimentally tested.

Advances in experimental systems~\autocite{mathys2024single, liu2025single, xiang2024gut, o2025genomics} now make it possible to evaluate AI predictions across the scales at which therapies must act~\autocite{qiu2022multimodal, andrieu2025harnessing}. Single-nucleus and single-cell atlases map millions of neurons and glia in the adult human brain, providing cell-type context for neurological disease~\autocite{siletti_transcriptomic_2023, kamath_single-cell_2022}. Genome-wide perturbation screens and protein-protein interaction assays reveal molecular mechanisms~\autocite{chung_identification_2013, mittal2017beta2}, while environmental toxicology~\autocite{paul_pesticide_2023}, deep proteomics~\autocite{bai2020deep}, and large clinical datasets~\autocite{orru2025diagnostic, sun2023human, llibre2025longitudinal} extend analyses from cellular to patient phenotypes. Each of these data scales captures a different dimension of brain disease biology, but these datasets are typically used in isolation. Graph-based AI models can reason over relations that link these dataset scales~\autocite{li2022graph, huang_foundation_2024}, but integration alone does not show whether AI predictions are biologically and clinically meaningful. The difficulty lies in the complexity of disease mechanisms~\autocite{bonev2024opportunities, rhinn2013integrative}: computational benchmarks based on retrospective label recovery can inflate performance through information leakage or task-specific artifacts~\autocite{wang2023scientific, ektefaie2024evaluating}, and in brain disease, this limitation is pronounced because mechanisms are distributed across genes, cells, and patient phenotypes~\autocite{gan2018converging,wang2025multiscale}.

To generate AI predictions that withstand validation across molecular, cellular, and clinical systems in neurology, we developed \model, a \underline{p}re-trained \underline{r}elati\underline{o}nal \underline{t}ransf\underline{o}rmer for \underline{n}eurology (Figure~\ref{fig:overview}). \model is a 578-million-parameter relational transformer~\cite{hu_heterogeneous_2020} trained on \kg, a knowledge graph contextualized to the adult human brain. Transformers are model architectures that use self-attention and next token prediction objectives~\cite{vaswani_attention_2017} and underpin many large-scale foundation models~\cite{wang2023scientific}. In biology, transformer models have been trained on genomic and protein sequences to learn representations that capture structure and function~\cite{nguyen2024sequence,lin2023evolutionary}, and on molecular structures to predict interactions~\cite{abramson2024accurate}. Transformers trained on single-cell expression datasets learn latent states and cell-type programs~\cite{cui2024scgpt}, and transformers trained on longitudinal personal health data learn the natural history of human disease~\cite{waxler2025generative,jiang2023health}. \model is a transformer architecture trained on a multimodal, multi-scale graph that links genes, proteins, cell types, brain regions, environmental exposures, phenotypes, and drugs, and performs relational reasoning across biological levels in neurological disease.

We used \model to generate hypotheses and tested them for three neurological diseases: Parkinson's disease (PD), bipolar disorder (BD), and Alzheimer's disease (AD), each evaluated in an independent biological system (Figure~\ref{fig:overview}a). In PD, \model reproduced the findings of six genome-scale $\alpha$-synuclein studies and linked genome-wide association loci to genes essential for dopaminergic neuron survival. After few-shot adaptation, \model also prioritized pesticides toxic to patient-derived dopaminergic neurons. In BD, \model nominated a compound that restored disrupted proteomic programs in cortical organoids from patients, producing effects distinct from and partly overlapping with lithium, the current frontline treatment for BD. In AD, \model predicted drugs associated with reduced seven-year dementia risk in survival analyses on 610,524 patients. \model generates hypotheses that we validated in molecular, organoid, and clinical systems, defining a path for AI-driven discovery in neurology.

%% file: 020results.tex
\subsection*{Development of \model}\label{sec:model-overview}

\model is a 578-million-parameter relational transformer for neurological disease (Figure~\ref{fig:overview}c). It uses a heterogeneous graph transformer architecture~\cite{hu_heterogeneous_2020} to learn the relationships among genes, proteins, cell types, brain regions, phenotypes, and drugs in \kg  (Methods Sec.~\ref{methods:model-architecture}). Unlike language models that read one token after another in a sequence~\cite{naveed2025comprehensive}, \model works on a multimodal biological network~\cite{zitnik2024current}, so it can reason over how genes, proteins, cell types, and drugs are connected. \model was trained on the \kg dataset using a self-supervised link prediction objective (Methods Sec.~\ref{methods:self-supervised-pre-training}). Through Bayesian hyperparameter optimization~\cite{victoria2021automatic}, we selected a model architecture that achieved high link prediction performance (AUROC $=0.9145$; accuracy $=82.23\%$) on the held-out test set.

\model produces an embedding for every node in the \kg knowledge graph. These embeddings organize nodes by shared biology, and we evaluated this organization by testing whether related nodes cluster together in the embedding space. Nodes cluster according to shared functional and clinical properties (Supplementary Note~\ref{si:note:emb-biomedically-organized}, Supplementary Figure~\ref{si:fig:umap_organization}). Diseases cluster by clinical similarity (neurodegenerative disease silhouette score $=0.1847$; Supplementary Note~\ref{si:note:diseases-cluster-latent}, Supplementary Figure~\ref{si:fig:disease_embeddings}), and drugs group by chemical structure (Supplementary Note~\ref{si:note:drugs-cluster-latent}, Supplementary Figure~\ref{si:fig:drug_embeddings}), showing that \model learns aligned relations across biological scales.

\model is trained on \kg, a multimodal biological network dataset contextualized to the human brain (Figure~\ref{fig:overview}b, Methods Sec.~\ref{method:building-kg}). \kg unifies 36 human datasets and ontologies (Methods Sec.~\ref{method:unifying-datasets}). We also integrated single-nucleus RNA-sequencing (snRNA-seq) atlases comprising 3,756,702 cells from the adult human brain, including 2,480,956 neurons, 888,263 non-neuronal cells, and 387,483 nuclei from PD patients and matched controls (Methods Sec.~\ref{method:contexualize-kg-brain}, Supplementary Table~\ref{si:table:siletti_cell_counts}). We identified differentially expressed and marker genes, validated them against immunohistochemistry of postmortem human brain, and used these data to define cell-type-specific nodes and edges (Supplementary Figure~\ref{si:fig:siletti}). \kg contains 147,020 nodes across 16 node types and 7,366,745 edges of 47 edge types (Supplementary Table~\ref{si:table:neurokg_nodes}, Supplementary Figure~\ref{si:fig:neurokg_stats}). We confirmed edge validity using PaperQA2, a generative literature search agent that has been validated to perform at or above the level of human experts on literature search \autocite{skarlinski_language_2024, lala_paperqa_2023}. PaperQA2 identified published evidence supporting 83.87\% of edges within the neighborhoods of six major neurological diseases (Supplementary Note~\ref{si:note:ai-agent-validates-kg}, Supplementary Figure~\ref{si:fig:pqa_rating}).

\subsection*{\model's {\em in silico} screens predict genome-wide experimental hits}\label{sec:predicts-asyn-results}

Parkinson's disease (PD) is the second most common neurodegenerative disease worldwide~\cite{dorsey_global_2018}. In 2021, 11.77 million people lived with PD \autocite{luo_global_2025}, and this number is expected to rise to 25.2 million by 2050 \autocite{su_projections_2025}. Although the defining features of PD are loss of dopaminergic (DA) neurons in the substantia nigra pars compacta and intracellular accumulation of the $\alpha$-synuclein protein \autocite{dickson_neuropathological_2009}, much remains unknown about the molecular biology and genetics of this debilitating disease. Genome-wide screens have identified genes and proteins linked to $\alpha$-synuclein. We therefore asked whether \model could act as a computational proxy by comparing its predictions with six genome-wide studies of $\alpha$-synuclein biology.

\model was used to perform three genome-wide \textit{in silico} screens (Figure~\ref{fig:pd}a). First, \model ranked all protein-coding genes by the predicted likelihood of a genetic association with PD (blue bars in Figure~\ref{fig:pd}c). Next, \model ranked all proteins by the predicted likelihood of a protein-protein (PPI) with $\alpha$-synuclein (red bars in Figure~\ref{fig:pd}c). Finally, \model combined both screens by averaging these two likelihoods and re-ranking all proteins (``PD $+$ $\alpha$-synuclein'' screen, purple bars in Figure~\ref{fig:pd}c). In all three screens, a higher rank corresponds to a stronger \textit{in silico} prediction. Each screen used a specific line of evidence in \kg: genetic relationships with PD were predicted from gene-disease associations in population genetics studies, and protein-protein relationships with $\alpha$-synuclein were predicted from human PPIs (Methods Sec.~\ref{method:unifying-datasets}).

We then evaluated \model's \textit{in silico} screens against six genome-wide $\alpha$-synuclein screens, including a split-ubiquitin yeast two-hybrid system (MYTH), an ascorbate peroxidase proximity labeling assay (APEX2), and a high-depth targeted exome sequencing study (TES) in 496 synucleinopathy patients (Figure~\ref{fig:pd}b, Methods Sec.~\ref{methods:asyn-studies}). For each \textit{in silico} screen, we evaluated the enrichment of the experimental hits among \model's predicted {\em in silico} hits via gene set enrichment analysis (GSEA) \autocite{subramanian_gene_2005, mootha_pgc-1-responsive_2003}. To prevent information leakage, gene or protein nodes with a link to either the PD node or to the $\alpha$-synuclein node in \kg were excluded from the analysis. Overrepresentation was quantified via a normalized enrichment score (NES), a Kolmogorov-Smirnov statistic adjusted for multiple hypothesis testing and gene set size. We compared \model performance against a random walk with restart (RWR) algorithm \autocite{tong_fast_2006, kohler_walking_2008}, a network science method \autocite{le_random_2017} that has enabled the discovery of genetic modules \autocite{cowen_network_2017}, drug targets \autocite{lee_identification_2018}, and disease-associated pathways \autocite{ghulam_disease-pathway_2020}. Starting at the PD node, RWR performed 10,000 random walks of length 10 in \kg. Protein nodes were then ranked by visitation frequency to predict a list of PD-associated proteins (gray bars in Figure \ref{fig:pd}c).

\model's combined PD $+$ $\alpha$-synuclein screen significantly enriched for experimental hits in five of six experimental screens, including in the MYTH (NES $= 2.30$, FDR-adjusted $p$-value $< 1 \times 10^{-4}$), APEX2 (NES $= 2.16$, FDR $< 1 \times 10^{-4}$), and TES (NES $= 2.13$, FDR $< 1 \times 10^{-4}$) screens (Figure~\ref{fig:pd}c). Moreover, the enrichment scores of the combined screen exceeded those of the RWR baseline in all six experimental screens, including by 74\% in MYTH, 57\% in APEX2, and 59\% in TES. \model's two non-combined \textit{in silico} screens (the PD association screen and the $\alpha$-synuclein interaction screen) were also informative but generally weaker than the combined PD + $\alpha$-synuclein screen. The $\alpha$-synuclein interaction screen alone was significant for all six experimental screens (for example, MYTH: NES = 2.09, FDR = $1.01 \times 10^{-3}$), whereas the PD association screen alone was significant for five of six (for example, MYTH: NES = 1.72, FDR = $5.02 \times 10^{-3}$; the mass spectrometry immunoprecipitation screen was not significant). In five of six experimental screens, the strongest enrichment was obtained by the combined PD $+$ $\alpha$-synuclein screen.

\subsection*{\model links Parkinson's genetics and molecular data}\label{sec:eval-gwas-essentiality}

We next examined whether \model can predict links between PD genetic risk loci from population studies and molecular hits identified in experimental data. Only 5-10\% of people with PD have familial forms explained by high-penetrance causal variants (\textit{SNCA}, \textit{LRRK2}, \textit{VPS35}, \textit{PRKN}, \textit{PINK1}, \textit{DJ1})~\cite{morris_pathogenesis_2024}, whereas genome-wide association studies (GWAS) implicate hundreds of loci of modest effect~\cite{simon-sanchez_genome-wide_2009,nalls_identification_2019}. How these risk alleles give rise to the neurodegenerative processes observed in PD remains unclear. We asked whether \model could predict links between genetic risk loci and molecular mechanisms, including those acting in vulnerable DA neuronal populations.

Drawing upon population genetic studies, including GWAS, rare-variant association studies, expression quantitative trait loci, and Mendelian randomization experiments (Figure \ref{fig:pd}d) \autocite{nalls_identification_2019, bustos_whole-exome_2020, foo_identification_2020, liu_genome-wide_2021, kamath_single-cell_2022, makarious_large-scale_2023, park_ethnicity-_2023, senkevich_association_2023, hop_systematic_2024, kim_multi-ancestry_2024, nazeen_deep_2024, nazeen_nerine_2025}, we considered 288 genes linked to PD.  For comparison, we also queried the NHGRI-EBI GWAS Catalog database \autocite{cerezo_nhgri-ebi_2025} to retrieve 8,976 genes linked to 10 other diseases across 863 GWAS studies. We also considered 681 genes that were found to be essential for the survival of DA neurons in an unbiased whole-genome CRISPR screen \autocite{nazeen_nerine_2025} (Figure \ref{fig:pd}e, Methods Sec.~\ref{methods:essentiality-screen}). For each DA-essential gene, \model performed an \textit{in silico} proteome-wide screen by predicting the likelihood of protein-protein interactions with all other proteins, resulting in a ranked list of proteins based on their likelihood of interaction with the DA-essential gene. Then, for PD and 10 other diseases, we identified the positions of each disease's GWAS hits via median rank, producing a distribution of ranked essential genes per disease. Known interactors with essential genes in the human reference interactome were excluded from the analysis. DA-essential genes were also removed from GWAS lists to prevent tautological enrichment.

\model predicted the strongest links between essential genes and GWAS hits from PD (median-median rank $=$ 9,487.5; Kruksal-Wallis $H =$ 1,942.62, $p < 0.001$) versus 10 other diseases: inflammatory bowel disease (10,020), asthma (10,276.5), type 2 diabetes mellitus (10,321.5), AD (10,523.5), coronary artery disease (10,686.5), breast cancer (10,698.5), hypertension (10,985), major depressive disorder (11,281), amyotrophic lateral sclerosis (11,441.5), and chronic kidney disease (11,601) (Figure \ref{fig:pd}f). Pairwise one-sided Mann-Whitney $U$ tests of PD versus each comparator disease were also significant (FDR-adjusted $p$-value $< 0.001$). Although diseases differed in the number of GWAS hits, these findings were robust to list-size differences: after randomly subsampling 150 GWAS genes per disease to equalize list sizes, PD still yielded the highest median ranks. Thus, \model links genes essential to the survival of dopaminergic neurons to PD risk genes more strongly than to GWAS hits from other diseases.

Next, we tested whether \model's predictions align with those of NERINE, a statistical framework for network-based rare variant association testing \autocite{nazeen_nerine_2025}. From the set of DA-essential genes, NERINE identified a module of 14 DA-essential genes enriched in autophagy regulation with a significant rare variant burden in patients with sporadic PD compared to aged controls in the Accelerating Medicines Partnership for Parkinson's disease program (AMP-PD)~\cite{iwaki2021accelerating} and UK Biobank cohorts. None of these genes were previously implicated in common variant GWAS of PD. We asked whether \model could preferentially associate known PD GWAS loci with these essential genes. For each of the 14 genes, we used \model to perform an \textit{in silico} genome-wide screen; then, we recorded the minimum rank achieved by any PD GWAS gene for each module member and contrasted this value against an empirical null obtained from 100 random draws of 288 non-PD genes. The predicted minimum PD rank fell below the corresponding null mean $\pm$ 95\% CI for 8 of the 14 genes, and exceeded it for only 3 of the 14 genes (\textit{TAB2}, \textit{OSBPL7}, and \textit{DRAM2}) (Figure \ref{fig:pd}g); of these, NERINE also predicts \textit{DRAM2} to have no effect on sporadic PD risk within the module. Further, two autophagy genes were nominally significant for \model-predicted interaction enrichment with PD GWAS genes: \textit{EXOC4} (minimum PD rank $= 3$, empirical one-sided $p$-value $= 0.05$) and \textit{HAX1} (minimum PD rank $= 4$, empirical one-sided $p$-value $= 0.03$), and HAX-1 has been reported as a component of brain stem and cortical Lewy bodies in PD and dementia with Lewy bodies \autocite{kawamoto_accumulation_2020}. 

\subsection*{\model predicts pesticides toxic to Parkinson's patient neurons}\label{sec:pd-pesticide-prediction}

The limited heritability of PD suggests that, beyond genetics, environmental factors also contribute to disease risk. Epidemiological studies have linked pesticide or pollutant exposure with elevated PD risk~\cite{petrovitch_plantation_2002,baldi_association_2003,ascherio_pesticide_2006,tanner_rotenone_2011,zhang_lewy_2025}, and some have suggested that the global rise in PD prevalence may be driven in part by increased pesticide exposure~\cite{dorsey_parkinsons_2024}. Nonetheless, most pesticides remain understudied in relation to PD, with robust evidence limited to pesticides such as paraquat and rotenone~\cite{liu_scientometric_2020}.

We fine-tuned \model to predict pesticides that increase risk for PD. For fine-tuning, we first curated a dataset of 28 pesticides that were associated with increased PD risk in a pesticide-wide association study (PWAS)~\cite{paul_pesticide_2023} conducted on PD patients ($n = 829$) and controls ($n = 824$) in the Parkinson's Environment and Genes study~\cite{ritz_pesticides_2016} (Figure~\ref{fig:pd}h, Methods Sec.~\ref{methods:pd-pwas}). The fine-tuning dataset comprised these 28 PWAS hits (positive samples) and 100 randomly sampled pesticides (negative samples). We fine-tuned \model to predict whether a given pesticide was a PWAS hit (Figure~\ref{fig:pd}j, Methods Sec.~\ref{methods:pesticide-ft}). After fine-tuning, we used \model to perform an \textit{in silico} pesticide-wide screen ($n = 696$ pesticides) by predicting which pesticides were most likely to increase risk for PD. To evaluate \model's performance, we used a held-out test set consisting of four PWAS hits that were shown to be toxic to iPSC-derived midbrain DA neurons from a PD patient overexpressing wild-type $\alpha$-synuclein~\autocite{paul_pesticide_2023} (Figure~\ref{fig:pd}i).

\model ranked the pesticides that were toxic to dopamine neurons near the top of its prediction list: three of the four test pesticides fell within the top quartile of all 696 predictions. \model predicted the organochlorine insecticide endosulfan at position 9 (top 1.29\%), the organochlorine insecticide dicofol at position 118 (top 16.95\%), and the organophosphate insecticide Naled at position 119 (top 17.09\%) (Figure~\ref{fig:pd}k). These results indicate that, despite being fine-tuned on a small dataset (28 positive and 100 negative pesticides), \model can predict pesticides that not only increase PD risk but are also directly toxic to dopaminergic neurons \textit{in vitro}.

\subsection*{\model forecasts repurposed drugs in neurological disease}\label{sec:disease-splits}

We next turned from predicting toxic compounds to predicting effective drugs for neurological disease. We evaluated \model for drug repurposing~\cite{pushpakom_drug_2019,tanoli_computational_2025}. Drugs whose targets are supported by human genetics are more than twice as likely to be approved~\cite{nelson_support_2015,minikel_refining_2024}, and many repurposing strategies search for drugs that act on genes with genetic links to disease~\cite{rodriguez_machine_2021,wu_integrating_2022,shuey_genetically_2023}. Instead of matching drugs to diseases by simple overlaps in these lists, \model uses its graph transformer to aggregate information along paths that connect drugs and diseases in \kg and scores each drug-disease pair based on this network context, capturing both direct and indirect therapeutic signals. To do so, \model draws on 28,774 drug-gene and drug-protein relationships (targets, transporters, metabolizing enzymes), 63,340 drug-disease links (indications, contraindications, off-label use, clinical trials), 64,249 drug-phenotype relationships, and 1,433,261 drug-drug interactions (Methods Sec.~\ref{method:unifying-datasets}, Supplementary Figure~\ref{si:fig:neurokg_stats}b).

We benchmarked \model's ability to forecast therapeutic approvals for 17 neurological diseases. These included common central and peripheral nervous system disorders from the 2021 Global Burden of Disease study~\cite{steinmetz_global_2024}, mental disorders from the 2019 GBD study~\cite{gbd_2019_mental_disorders_collaborators_global_2022}, and selected rare neurological diseases (\eg, X-linked dystonia-parkinsonism). For each disease, we asked whether \model could predict drugs that are FDA approved, prescribed off-label, or supported by strong clinical evidence (Methods Sec.~\ref{method:unifying-datasets}).

As many of these approvals and evidence signals are already captured in the \kg pre-training data via sources like DrugBank \autocite{wishart_drugbank_2018}, Drug Central \autocite{avram_drugcentral_2021}, and the Open Targets Platform \autocite{ochoa_next-generation_2023, koscielny_open_2017}, we introduce disease-centric data splits to evaluate model generalizability  and prevent information leakage (Figure~\ref{fig:bd}a). For each disease in the evaluation set, we removed all drug-disease edges involving that disease as well as its related diseases from \kg before training (Methods Sec.~\ref{methods:disease-splits}). \model was trained from scratch on each disease-specific data split of \kg, and we evaluated each model's ability to predict drugs that are FDA approved, in late-stage clinical trials, or prescribed off-label for the specific disease(s). For each disease, \model was used to conduct an \textit{in silico} screen over all 8,160 drugs in \kg using the ``indication'' edge type. We quantified performance using macro-averaged recall ($R$) at $k$, defined as the average percentage of known drugs that \model recovers for a disease at various ranking thresholds $k$. Specifically, we measured recall at $k = 408$ drugs (top 5\%, $R = 47.87\% \pm 30.99\%$, mean $\pm$ SD), $k = 816$ drugs (top 10\%, $R = 75.69\% \pm 22.61\%$), $k = 1,224$ drugs (top 15\%, $R = 89.24\% \pm 11.74\%$), $k = 1,632$ drugs (top 20\%, $R = 93.34\% \pm 8.30\%$), and $k = 2,040$ drugs (top 25\%, $R = 96.24\% \pm 5.47\%$) (Figure~\ref{fig:bd}b). For 6 and 13 diseases, \model predicted over 80\% of all drugs among the top 10\% and 15\% of its predictions, respectively, suggesting that \model can narrow the drug repurposing search space in neurological disease.

\subsection*{\model predicts a drug that restores bipolar organoid proteomes}\label{sec:bd-organoids}

We used \model to predict drugs for bipolar disorder (BD), a life-long neuropsychiatric disorder with unmet clinical need characterized by recurring manic and depressive episodes. With an aggregate lifetime prevalence of 2.4\%~\cite{merikangas_prevalence_2011}, BD affects over 40 million people worldwide~\cite{singh_bipolar_2025}. Since the late 1990s, lithium has remained the first-line treatment for BD~\cite{geddes_treatment_2013}; however, even when lithium is administered, approximately 37\% of patients relapse within 1 year, 60\% within 2 years, and up to 87\% after 5 years~\cite{gitlin_relapse_1995,keller_bipolar_1993}. Antidepressants have also been used to treat depressive episodes in BD, but evidence for their efficacy is limited, and current treatment guidelines recommend against their use~\cite{vieta_antidepressants_2014,pacchiarotti_international_2013}. These limitations of BD treatments motivate complementary strategies such as drug repurposing.

We used \model to predict repurposing candidates for BD. First, we trained \model on a BD-specific data split and measured recall at various values of $k$ for drugs directly indicated or prescribed off-label for BD. \model achieved recall of 81.25\% in the top 10\% and 93.75\% in the top 20\% of predicted drug indications. For off-label BD drugs, recall reached 94.73\% in the top 10\% and 100\% in the top 25\%, indicating that \model predicted known BD therapies even when trained without access to any BD-drug information. We then re-trained \model on the full \kg and used the resulting rankings to select candidate drugs for experimental testing in cortical organoids derived from patients with BD.

A challenge in translating candidate drugs for BD is that no experimental model fully captures the clinical syndrome or genetic heterogeneity seen in patients. Animal models are particularly limited: mouse models of BD often rely on psychostimulant-induced mania or continuous hyperactivity driven by mutations in the \textit{Clock} gene, even though mutations in circadian genes like \textit{Clock} are not present in most BD patients and the overlap between psychostimulant response and BD pathophysiology is poor~\autocite{kato_animal_2016, beyer_animal_2017}. Human brain organoids derived from patient cells offer a complementary \textit{in vitro} system that captures patient-specific genetic backgrounds and cell-type programs, although they also do not recapitulate the full clinical syndrome~\autocite{quadrato_promises_2016}. To test candidate drugs predicted by \model in a disease-relevant human context, we paired \model with a cortical organoid model of BD~\autocite{meyer_impaired_2024}. We considered the top 200 drugs predicted by \model for BD and used an LLM evaluator followed by expert review to highlight notable drug candidates (Figure~\ref{fig:bd}c, Methods Sec.~\ref{methods:bd-candidate-pred}). \model surfaced several compounds with little to no prior investigation in BD, including the three sequential vitamin D metabolites cholecalciferol (\#16), alfacalcidol (\#102), and calcitriol (\#114). Calcitriol, the active form of vitamin D, acts as a neurosteroid and has been implicated in brain development, neurotransmission, and neuroprotection via the nuclear vitamin D receptor (VDR)~\autocite{haussler_nuclear_1998}. Epidemiological studies associate vitamin D deficiency with increased risk of neuropsychiatric conditions, and supplementation has been reported to benefit mood symptoms, including in BD~\autocite{groves_vitamin_2014, harms_vitamin_2011, eyles_distribution_2005, cereda_role_2021}. We therefore treated  \textit{in vitro} BD cortical organoids~\autocite{meyer_impaired_2024} with calcitriol to investigate its potential neuroprotective, neuromodulatory, and neurometabolic effects.

We treated cortical organoids derived from $n = 5$ bipolar patients and $n = 4$ healthy controls (Supplementary Table~\ref{si:table:ipsc-lines-organoid}) with calcitriol for one week, then performed unbiased deep proteomic profiling using an Orbitrap Astral Mass Spectrometer (Figure~\ref{fig:bd}d, Methods Sec.~\ref{methods:organoids}). Proteomic analysis identified 512 proteins that were significantly differentially expressed between BD and control (CTR) cortical organoids (FDR-adjusted $p$-value $< 0.05$) (Figure~\ref{fig:bd}e). Comparison of untreated and calcitriol-treated BD organoids revealed 633 significantly differentially expressed proteins (FDR $< 0.05$; Figure~\ref{fig:bd}f). To elucidate the effects of calcitriol on the BD proteome, we examined the 73 proteins that overlapped between the BD~vs.~CTR and BD~vs.~BD~+~calcitriol datasets (Figure~\ref{fig:bd}g). Consistent with previous findings \autocite{meyer_impaired_2024}, untreated BD and CTR organoids formed distinct clusters after unsupervised hierarchical clustering analysis. Notably, 4 out of 5 calcitriol-treated BD samples clustered more closely with controls than with untreated BD organoids, indicating a substantial normalization of the BD proteome following calcitriol treatment.

Further analysis of these proteins highlighted key processes implicated in BD pathophysiology: ribosomal function and mRNA processing \autocite{darby_consistently_2016, sharma_ribosome_2024, salvetat_game_2022}, metabolic and protein homeostasis (including fructose and amino acid metabolism) \autocite{fagiolini_bipolar_2008}, and synaptic adhesion and plasticity (Figure~\ref{fig:bd}h). Specifically, studies have shown an association between BD and RNA alterations by epitranscriptomic mechanisms \autocite{pisanu_understanding_2018}, including RNA methylation, microRNAs \autocite{zhao_genome-wide_2015}, and RNA editing \autocite{salvetat_game_2022}. Among these RNA processes, the adenosine (A)-to-inosine (I) conversion is frequently studied; these conversions are mediated by ADARs (Adenosine Deaminase Acting on RNA), which bind to double-stranded RNA stem loops and modify A to I by deamination. It was previously shown that the RNA editome could be used to differentiate BD from unipolar, strongly implicating ADAR dysregulation in BD \autocite{salvetat_game_2022}. Notably, calcitriol treatment reversed ADAR overexpression in our cortical organoid model, suggesting a potential therapeutic role for vitamin D signaling in modulating RNA editing in BD (Figure~\ref{fig:bd}h). In addition, calcitriol treatment reversed expression of PDE6D, REV3L, and thymosin $\beta_4$ proteins, whose corresponding genes have also been shown to be dysregulated in genome-wide expression studies of the postmortem bipolar brain \autocite{nakatani_genome-wide_2006}. Lastly, we also conducted comparative proteomic analyses of BD cortical organoids following treatment with calcitriol and lithium orotate (Figure~\ref{fig:bd}i). Each drug affected distinct clusters of proteins, with only partial overlap between the two, suggesting that calcitriol and lithium act through substantially different mechanisms of action. 

\subsection*{\model predicts drugs linked to lower Alzheimer's dementia risk}\label{sec:ad-clinical-trial}

We next sought to evaluate \model-predicted drugs in real-world patient data. We applied \model to Alzheimer's disease and related dementias (ADRD) using retrospective analyses of longitudinal personal health data from 610,524 patients in the Mass General Brigham (MGB) healthcare system. We evaluated whether candidate drugs predicted by \model were associated with reduced ADRD risk (Figure~\ref{fig:ad}a). Candidate drugs were selected through a multi-stage filtering process. First, \model performed an \textit{in silico} screen to rank all 8,160 drugs in \kg by predicted likelihood for ADRD. We then considered drugs with \model-predicted ranks $< 60$ that were FDA approved for non-dementia indications associated with specific ICD-10 diagnosis codes in the personal health data (Supplementary Table~\ref{si:table:icd10-diagnosis}); this step excluded, for example, antibiotics, sedatives, anticoagulants, and analgesics. Indications for each drug were determined by LLM query~\cite{openai_gpt-4o_2024} and manual expert review. Finally, to ensure sufficient statistical power, we retained only those drugs for which more than 1,000 patients at MGB received treatment, yielding $n = 8$ drugs predicted by \model.

For each drug, we constructed a retrospective cohort of individuals aged $\geq 50$ years who had the relevant primary disease indication, defined by the corresponding ICD-10 diagnosis code in MGB's electronic health record (EHR) system. Patients prescribed the candidate drug formed the treatment group; the control group comprised patients with the same indication who were not prescribed the candidate drug or any other drug ranked highly by \model for that indication. We used Cox proportional hazards models to estimate the association between drug exposure and incident dementia, comparing ADRD-free survival between treatment and control groups. Study entry was the date of indication diagnosis for controls and the first prescription date for treated patients. Participants were followed until the first dementia diagnosis (any of 22 ADRD-associated ICD-10 codes, which were curated by expert neurologists) or were censored at death or at last encounter within the MGB healthcare system. To address confounding by indication, we used inverse probability treatment weighting (IPTW) to balance treatment and control groups on age at entry and sex, and also included these covariates in the doubly robust Cox regression model.

Five of the eight evaluated drugs were associated with a significantly reduced risk of ADRD (Figure~\ref{fig:ad}b-d). Total $n=610,524$ patients were included in the analysis (Figure~\ref{fig:ad}c). The strongest protective association was observed for aflibercept (indicated for wet age-related macular degeneration), which was linked to a 37\% reduction in dementia risk (hazard ratio [HR] $= 0.63$, 95\% CI $[0.53-0.75]$, $p$-value $= 1.1 \times 10^{-7}$; Figure~\ref{fig:ad}e). The SGLT-2 inhibitor dapagliflozin (indicated for type 2 diabetes) was also strongly associated with a lower risk (HR $= 0.70$, 95\% CI $[0.57-0.86]$, $p = 7.9 \times 10^{-4}$; Figure~\ref{fig:ad}f). Among the antihypertensives, valsartan (HR $= 0.83$, 95\% CI $[0.79-0.88]$, $p = 1.2 \times 10^{-10}$; Figure~\ref{fig:ad}g), spironolactone (HR $= 0.89$, 95\% CI $[0.84-0.95]$, $p = 1.8 \times 10^{-4}$), and metoprolol (HR $= 0.96$, 95\% CI $[0.93-0.99]$, $p = 0.004$) all showed significant protective associations. The remaining three drugs (antihypertensive ramipril and cholesterol-lowering medications rosuvastatin and ezetimibe) did not significantly increase or decrease ADRD risk.

%% file: 030discussion.tex
Neurological diseases are the leading cause of disability and the second leading cause of death worldwide, yet for most, there are no treatments that slow or halt disease progression. We developed \model, a graph AI model that generates hypotheses for neurological disease by operating on \kg, a brain-centered multimodal knowledge graph contextualized with snRNA-seq of the adult human brain. We used \model to generate predictions in Parkinson's disease, bipolar disorder, and Alzheimer's disease and related dementias, and tested these predictions in genome-scale molecular screens, patient-derived organoids, and longitudinal electronic health record analyses.

We paired \model with experiments and clinical analyses in PD, BD, and AD to test its predictions in diverse systems. In PD, \model linked genes essential for dopaminergic neuron survival to genome-wide association hits and produced ranked lists that were significantly enriched for hits from six genome-scale $\alpha$-synuclein studies in \textit{in silico} screens. It connected DA-essential genes to PD GWAS loci more strongly than to GWAS hits from other diseases and highlighted specific genes that bridge common and rare variant signals. After few-shot fine-tuning, \model also prioritized pesticides toxic to patient-derived dopaminergic neurons, including endosulfan, ranked in the top 1.29\% of predictions. In BD, \model predicted drug candidates that could reverse disease-associated molecular programs, and experiments in cortical organoids derived from patients showed that calcitriol partially restored the BD proteome. In AD, \model predicted eight drugs associated with reduced dementia risk, which we evaluated in EHR data from $n = 610{,}524$ patients; five drugs were linked to a lower seven-year risk of dementia.

\model is applied to three neurological diseases, and extending it to the full spectrum of neurological conditions will require broader data coverage and additional validation. Many conditions, particularly rare and understudied diseases, are sparsely represented in datasets, which limits the scope of current predictions. Despite efforts to harmonize biomedical data, \kg itself remains incomplete. Gaps may arise from publication bias, underreporting of negative results, or uneven research focus that inflates connectivity for well-studied genes or diseases. Because \model is trained on \kg, such biases can propagate into its predictions. Transformer-based models on graphs, including \model, can be sensitive to node degree and may assign higher confidence to well-connected nodes~\autocite{subramonian_theoretical_2024, liu_generalized_2023, tang_investigating_2020}. We mitigated this effect by upweighting rare edge types and optimizing across relation classes, yet predictions may still favor high-degree nodes. In biomedical contexts where interpretability is essential, this tendency can be useful, but the balance between reliability and novelty remains unresolved. Addressing these limitations will require continually updated knowledge graphs that evolve with new neurological data, support ongoing model retraining, and enable \model to generalize across a wider range of neurological diseases.

AI models trained on biomedical data can generate hypotheses that are testable across molecular, cellular, and clinical systems. Our results suggest that, in PD, BD, and AD, multimodal data reconciled by AI can reveal how distinct biological factors converge on disease mechanisms and neuroprotective pathways. More broadly, \model illustrates how AI can propose testable hypotheses and candidate interventions in neurology, and points toward AI-experiment discovery loops that connect molecular data, experimental systems, and patient outcomes in human brain disease.

%% file: 200figures.tex

\begin{figure}[!t]
  \centering
  \includegraphics[width=\linewidth]{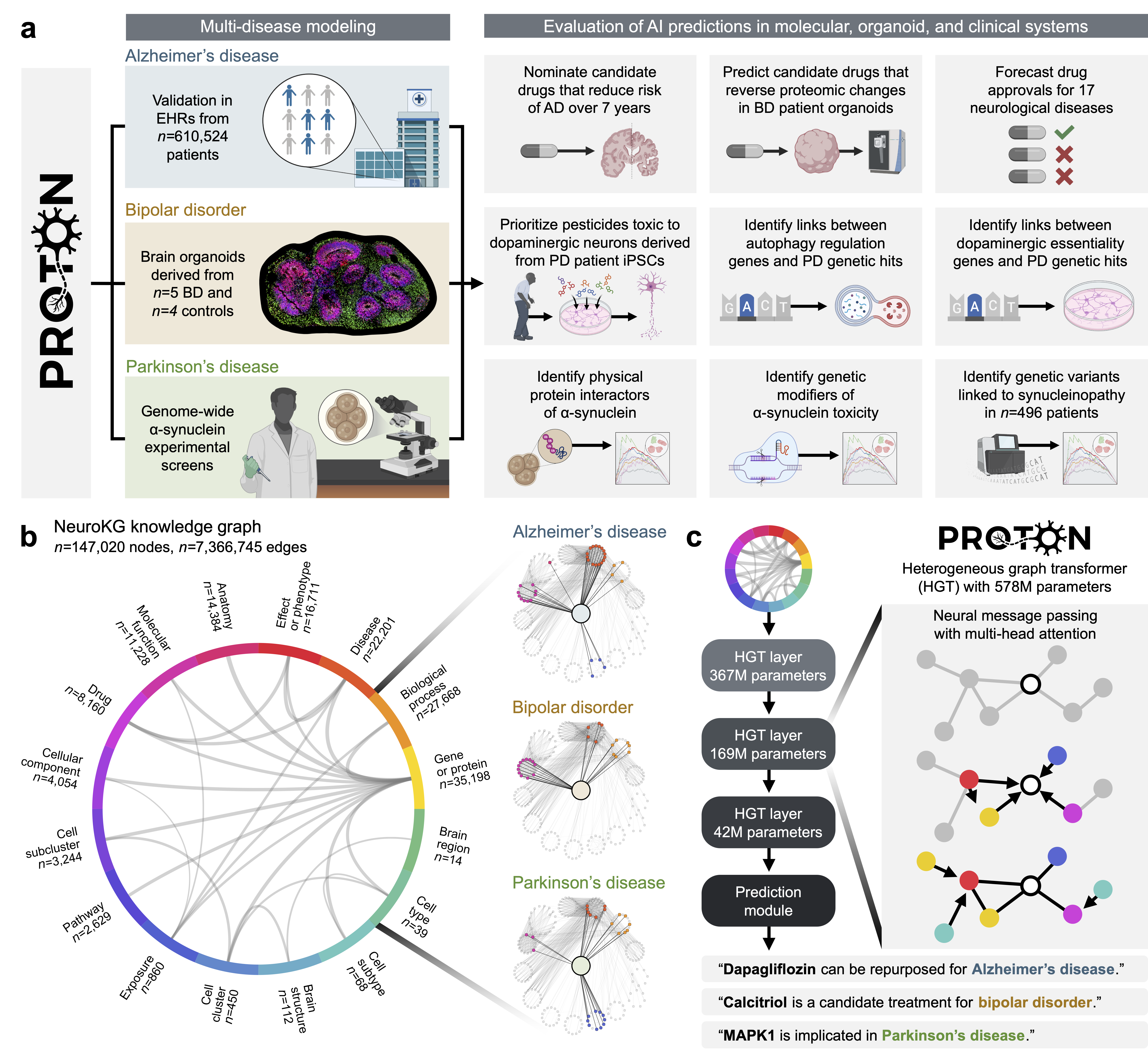}
  \caption{\textsbf{Overview of \model.} \model is a graph AI model for neurological disease.
  \textsbf{(a)} We demonstrate diverse disease-specific applications of \model using experimental and clinical data in three neurological conditions: Parkinson's disease (PD), bipolar disorder (BD), and Alzheimer's disease (AD). \model nominates candidate drugs, forecasts drug approvals, prioritizes pesticides, and identifies genetic, proteomic, and protein-protein interaction links across multiple biological scales.
  \textsbf{(b)} To develop \model, we first constructed \kg, a heterogeneous biomedical knowledge graph with 147,020 nodes (16 types) and 7,366,745 edges (47 types). Diseases have unique connectivity patterns in \kg; for example, 200 type-balanced neighbors from the two-hop neighborhoods of PD, BD, and AD are shown on a single shared layout. The one-hop neighborhood of each disease node is rendered in full color; two-hop and out-of-neighborhood nodes are rendered faintly. \textsbf{(c)} We pre-trained a 578-million-parameter heterogeneous graph transformer (HGT) on \kg with a self-supervised link-prediction objective. Node and edge type-aware multi-head attention enabled neighborhood integration across modalities.}
  \label{fig:overview}
\end{figure}
\clearpage

\begin{figure}[!t]
  \centering
  \includegraphics[width=\linewidth]{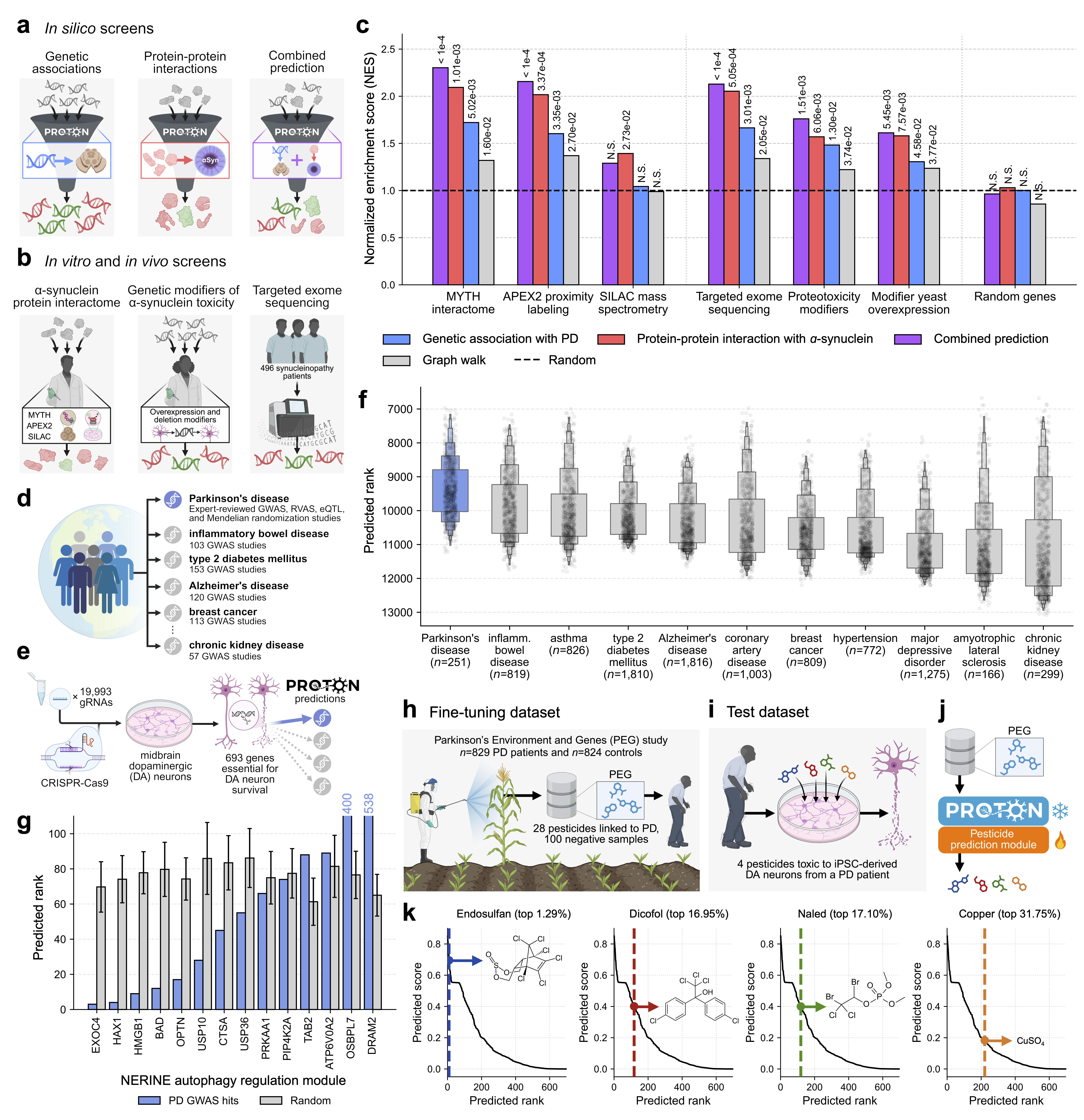}
  \caption{\textsbf{\model prioritizes genes and pesticides linked to Parkinson's disease.}
  \textsbf{(a)} \model conducted three proteome-scale \textit{in silico} screens: genetic relationship with PD, protein–protein interaction with $\alpha$-synuclein, and a combined prediction.
  \textsbf{(b)} \model predictions were compared to six genome-wide experimental $\alpha$-synuclein studies.
  \textbf{(c)} Gene set enrichment analysis (GSEA) demonstrated that \model predictions were significantly enriched for experimental hits, outperforming a graph random walk baseline. Results are grouped by experiment class: $\alpha$-synuclein interactome profiling or $\alpha$-synuclein genetic studies in humans or disease models. N.S. $=$ not significant. 
  \textsbf{(d)} We compiled 288 PD GWAS-linked genes across multiple genetic study types and contrasted them against 8,976 GWAS hits from 10 other diseases. 
  \textsbf{(e)} An unbiased whole-genome CRISPR screen in human midbrain dopaminergic (DA) neurons identified genes essential for neuronal survival. 
  \textsbf{(f)} For each essential gene, \model performed a proteome-wide \textit{in silico} screen to rank likely protein–protein interactors; the median rank of disease GWAS hits showed the strongest association for PD compared with other diseases.
  \textsbf{(g)} For 14 essential genes enriched in autophagy regulation and previously implicated in PD by NERINE, an independent network-based method for rare variant burden
  }
  \label{fig:pd}
\end{figure}
\clearpage

\noindent
\begin{figure}[H]\ContinuedFloat
  \caption*{
  analysis \autocite{nazeen_nerine_2025}, \model's minimum PD GWAS association rank per gene was compared to an empirical null, highlighting \textit{EXOC4} and \textit{HAX1}.
  \textsbf{(h)} Fine-tuning dataset constructed from the Parkinson’s Environment and Genes (PEG) study ($n=829$ PD patients and $n=824$ controls), comprising 28 pesticides linked to PD and 100 negative samples. 
  \textsbf{(i)} The held-out test dataset consisted of four pesticides toxic to iPSC-derived midbrain DA neurons from a PD patient.
  \textsbf{(j)} Schematic of the \model pesticide prediction pipeline.
  \textsbf{(k)} \model prioritized pesticides toxic to patient-derived dopaminergic neurons, including endosulfan, dicofol, Naled, and copper sulfate.
  }
\end{figure}

\begin{figure}[!t]
  \centering
  \includegraphics[width=\linewidth]{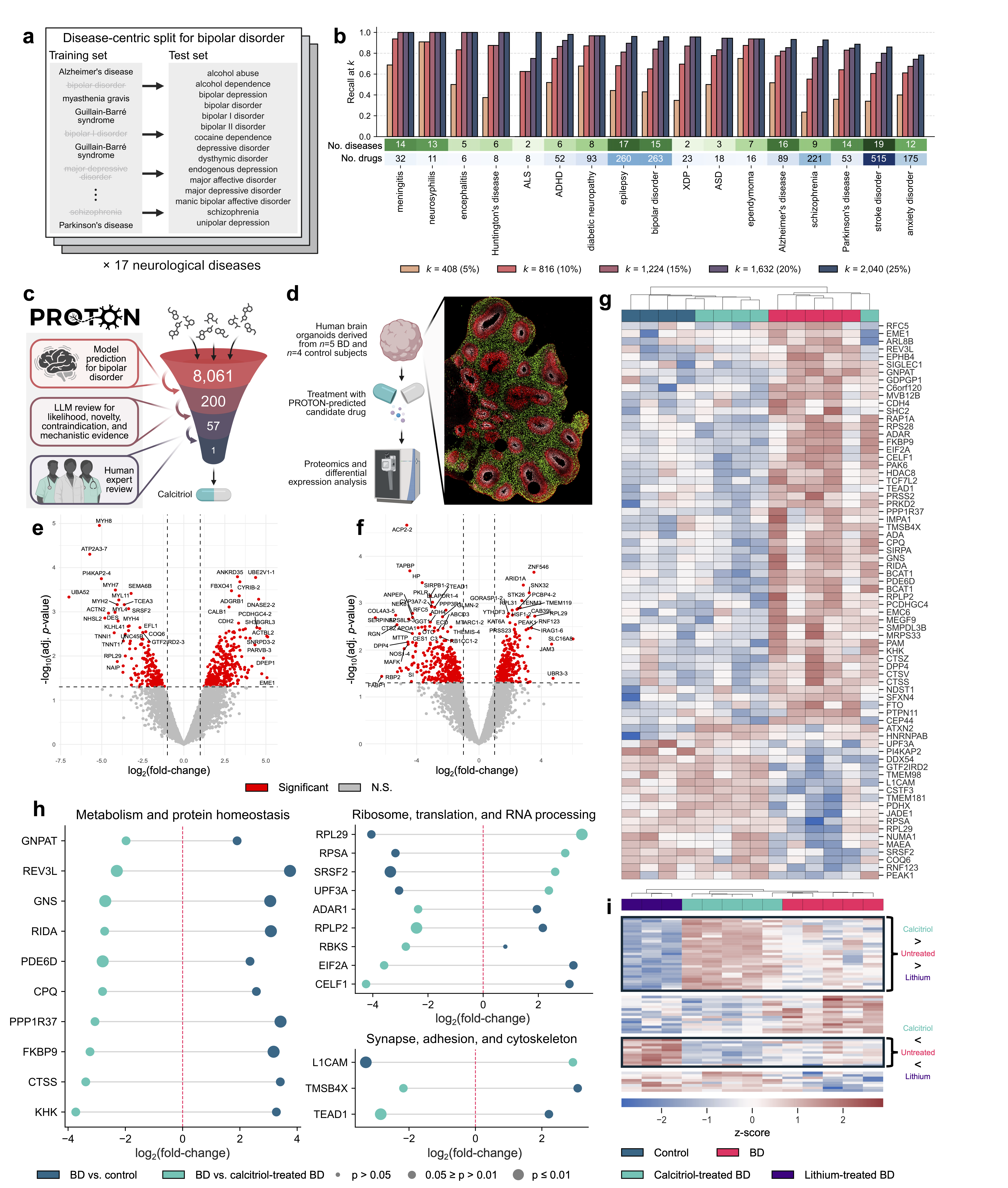}
  \caption{\textsbf{\model-nominated drug reverses proteomic alterations in patient-derived organoid models of bipolar disorder.}
  \textsbf{(a)} For each of 17 neurological diseases, we constructed disease-centric data splits by withholding drug information from the training set for the disease of interest and all related diseases. An example disease-centric data split is shown for BD.
  \textsbf{(b)} Across 17 neurological diseases, \model demonstrates strong performance at recovering drugs that are FDA approved, in late-stage clinical trials, or prescribed off-label, as measured 
  }
  \label{fig:bd}
\end{figure}
\clearpage

\noindent
\begin{figure}[H]\ContinuedFloat
  \caption*{
  by macro-averaged recall ($R$) at $k$.
  \textsbf{(c)} \model highlighted calcitriol as a candidate therapeutic for BD. 
  \textsbf{(d)} Human cortical organoids derived from BD patients and healthy controls were treated with calcitriol, followed by deep proteomic profiling. 
  \textsbf{(e–f)} Volcano plots show differentially expressed proteins (DEPs) in BD versus control organoids (512 DEPs, $p$ < 0.05) and in BD versus drug-treated BD organoids (633 DEPs, $p$ < 0.05). 
  \textsbf{(g)} Expression levels of 73 proteins dysregulated in BD organoids were normalized by calcitriol treatment, as demonstrated by the unsupervised clustering of calcitriol-treated BD organoids with control organoids versus untreated BD organoids.
  \textsbf{(h)} Key pathways affected by calcitriol treatment include metabolism and protein homeostasis, RNA processing, and synaptic adhesion. 
  \textsbf{(i)} Comparative proteomics revealed both distinct and overlapping proteomic effects of calcitriol and lithium in patient-derived organoids. \textit{Abbreviations:} attention deficit hyperactivity disorder (ADHD), amyotrophic lateral sclerosis (ALS), autism spectrum disorder (ASD), X-linked dystonia-parkinsonism (XDP).
  }
\end{figure}

\begin{figure}[!t]
  \centering
  \includegraphics[width=\linewidth]{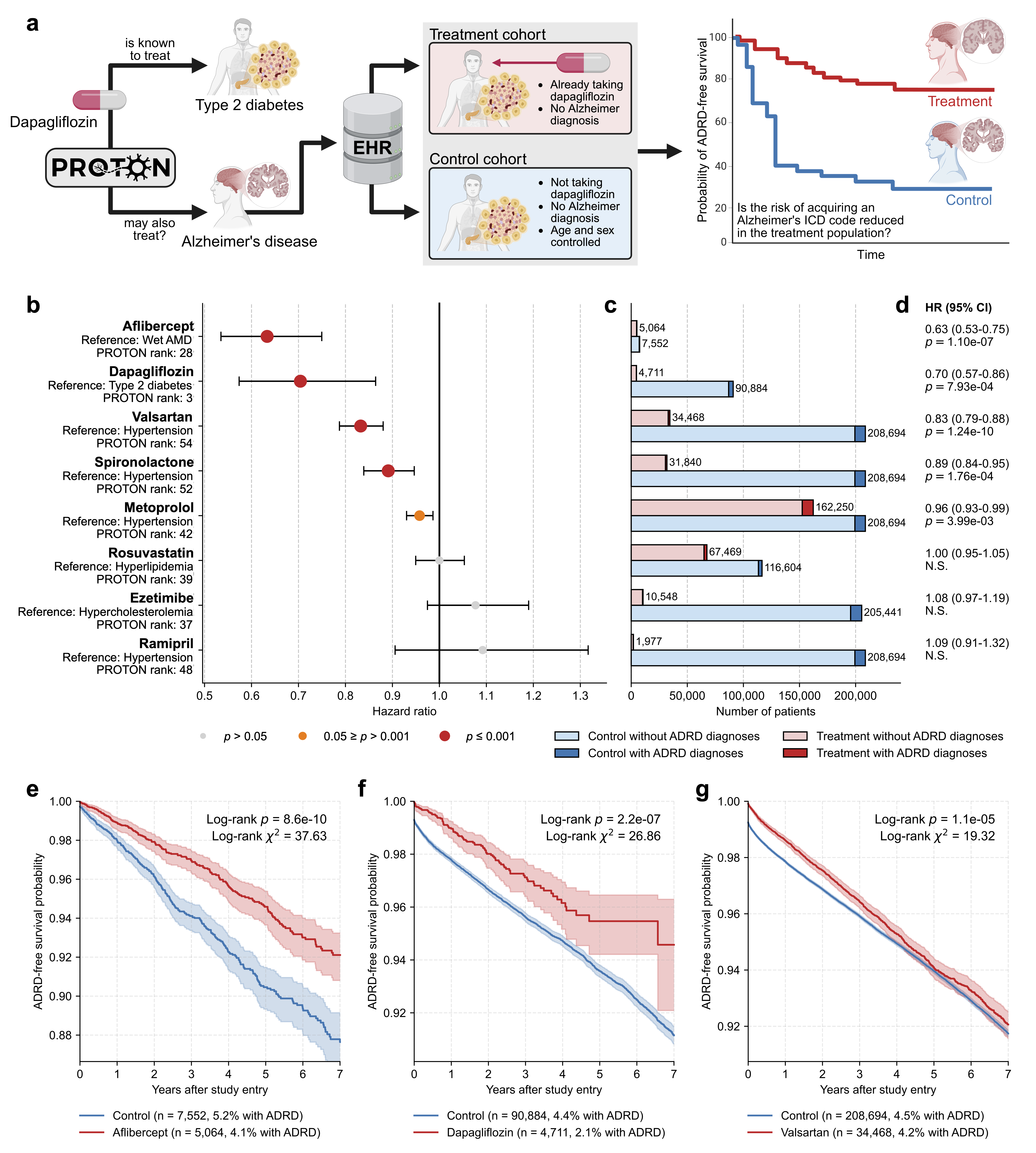}
  \caption{\textsbf{\model predicts drugs that reduce the risk of Alzheimer's disease and related dementias in EHR analyses of 610,524 patients.}
  \textsbf{(a)} To evaluate \model drug repurposing predictions, we conducted retrospective analyses of Mass General Brigham health records data across $n=610,524$ patients. For each candidate drug, patients with the relevant primary indication were divided into treatment and control cohorts. We compared ADRD-free survival between cohorts using Cox proportional hazards models with inverse probability treatment weighting for age and sex.
  \textsbf{(b)} Five of eight model-predicted drugs were significantly associated with reduced ADRD risk ($p < 0.05$). The other three drugs neither significantly increased nor decreased risk ($p > 0.05$).
  \textsbf{(c)} Cohort sizes for treatment and control groups across retrospective EHR analyses.
  \textsbf{(d)} Cox hazard ratios and $p$-values.
  \textsbf{(e–g)} Kaplan–Meier survival curves show reduced ADRD incidence for aflibercept, dapagliflozin, and valsartan. Log-rank $p$-values and $\chi^2$ statistics are reported.}
  \label{fig:ad}
\end{figure}

\clearpage

%% file: 100methods.tex
The Online Methods are organized into four sections. First, we describe the pre-training data, detailing the construction of the Neurological Disease Knowledge Graph (\kg) and its contextualization with human brain single-cell RNA-sequencing atlases (Section~\ref{method:pretraining-data}). Second, we outline the \model architecture and its self-supervised pre-training procedure (Section~\ref{methods:model-architecture-optimization}). Third, we describe the experimental studies conducted to generate independent datasets for evaluation (Section~\ref{methods:experimental-studies}). Finally, we specify the methods used to assess \model's performance on these datasets (Section~\ref{methods:evaluation-protocols}).

\section{Pre-training data}\label{method:pretraining-data}

\subsection{Building a heterogeneous biomedical knowledge graph}\label{method:building-kg}

Graphs are an attractive choice as a data model to describe the interacting elements in biomedical data, where networks are pervasive across multiple scales \cite{li_graph_2022}. For example, gene regulatory networks \cite{ravasi_atlas_2010}, protein-protein interaction networks \cite{luck_reference_2020}, cell-cell interactions, tissue spatial organization \cite{wang_deep_2023}, hierarchies of clinical knowledge, patient relationships in the electronic health record (EHR) \cite{hong_clinical_2021}, and other complex associations in the biological interactome can all be natively represented as graphs \cite{barabasi_network_2011}. In particular, such graphs with knowledge-informed node and edge types are often referred to as knowledge graphs (KGs). KGs are fundamental data structures represented by the tuple $\scriptG = (\scriptV, \scriptE)$, which is defined by a set of nodes $\scriptV$ and a set of edges $\scriptE$. Nodes in $\scriptV$ have different types $\scriptA$; for example, in biomedical KGs, nodes can represent proteins, diseases, cell types, or brain regions, among many other possibilities. Formally, let $\scriptA$ be the set of node types with mapping function $\phi: \scriptV \rightarrow \scriptA$ and let $\scriptR$ be the set of edge types. Then, the edges are represented as
$$\mathcal{E} \subseteq \{\{u, v\} \times \mathcal{R} \mid u, v \in \mathcal{V},\, u \neq v \},$$
where each edge is a pair $\bigl(\{u, v\}, r\bigr)$ consisting of an unordered set of nodes $\{u, v\}$, each belonging to a specific node type $\phi(u), \phi(v) \in \scriptA$, and connected to each other by an undirected edge of type \(r \in \mathcal{R}\). The classes, or types, of nodes and edges often indicate the nature of their relationships; for example, in a biomedical KG, nodes of type ``drug'' and ``disease'' may be connected by edges of type ``indication,'' ``contraindication,'' or other relation types.

Especially in biology and medicine, KGs have emerged as important data structures to represent and operate on complex data. Examples of biomedical KGs include the Precision Medicine Knowledge Graph (PrimeKG), a KG with 129,375 nodes and 4,050,249 edges created by integrating 20 primary source biomedical datasets, ontologies, and knowledge bases \cite{chandak_building_2023}; \href{https://ogb.stanford.edu/docs/linkprop/}{\texttt{ogbl-biokg}}, a heterogeneous KG from the Open Graph Benchmark compiled from various biorepositories with 93,773 nodes and 5,088,434	edges \cite{hu_open_2020}; \href{https://github.com/vaticle/biograkn}{BioGrakn}, a graph-based deductive database that includes a collection of biomedical knowledge graphs \cite{messina_biograkn_2018}; Hetionet, a heterogeneous biomedical KG created by integrated 29 public biomedical resources \cite{himmelstein_systematic_2017}; the \href{https://github.com/biocypher/clinical-knowledge-graph/}{Clinical Knowledge Graph (CKG)}, an open-source KG with 19,405,058 nodes and 217,341,612 edges that aims to capture clinically relevant information \cite{santos_knowledge_2022}; the \href{https://github.com/wcm-wanglab/iBKH}{integrative Biomedical Knowledge Hub (iBKH)}, constructed by merging 18 biomedical knowledge sources \cite{su_biomedical_2023}; PharmKG, a biomedical knowledge graph with 7,603 nodes and 500,958 edges for drug repurposing and target prediction \cite{zheng_pharmkg_2021}; \href{https://github.com/dsi-bdi/biokg}{BioKG}, a biomedical knowledge graph that draws from 14 biological databases \cite{walsh_biokg_2020}; the \href{https://github.com/gnn4dr/DRKG}{Drug Repurposing Knowledge Graph (DRKG)}, a KG with 97,238 nodes and 5,874,261 edges assembled from six databases and recent publications on COVID-19 \cite{ioannidis_drkg_2020}; and many other KGs. However, existing KGs suffer from various drawbacks: their indiscriminate ``kitchen sink'' construction methodologies often lead to noisy or incomplete graphs with erroneous or missing edges \autocite{zaveri_linked_2017, xue_knowledge_2022}. Further, no current KG exists that captures neurological and neuroscientific knowledge across multiple scales, including the molecular (\eg, protein-protein interactions), cellular (\eg, brain single-cell RNA-sequencing data), genetic, anatomical, and clinical levels. Thus, we sought to integrate diverse public information about neurobiological relationships into the \underline{Neuro}logical Disease \underline{K}nowledge \underline{G}raph (\kg), a harmonized data platform suitable for training AI models.

\subsection{Unifying 36 large-scale biomedical data sources}\label{method:unifying-datasets}

\kg is a heterogeneous biomedical knowledge graph created by unifying 36 high-quality, publicly available primary data sources. Each dataset was processed to construct both nodes and edges in \kg, where nodes represent objects in the datasets harmonized by mapping their various unique identifiers onto each other, and edges represent relationships between the objects similarly extracted from the primary data sources. Building on \textcite{chandak_building_2023}, we first analyzed and integrated data from 20 knowledge bases and ontologies, including human expression data from the Bgee gene expression knowledge base \autocite{bastian_bgee_2021}; environmental exposure data from the Comparative Toxicogenomics Database (CTD) \autocite{davis_comparative_2021}; the DisGeNET knowledge base of gene-disease associations \autocite{pinero_disgenet_2020}; the Disease Ontology database of human disease relationships \autocite{schriml_human_2019}; the DrugBank pharmaceutical database \autocite{wishart_drugbank_2018}; the Drug Central database of drug-disease interactions \autocite{avram_drugcentral_2021}; the Entrez gene database of gene-specific information \autocite{maglott_entrez_2011}; the Gene Ontology (GO) pathway database, from which biological processes, cellular components, and molecular functions were all included \autocite{the_gene_ontology_consortium_gene_2023, ashburner_gene_2000}; the Human Phenotype Ontology (HPO) \autocite{gargano_human_2024}; the Mayo Clinic knowledge base of information on human diseases and conditions; the MONDO Disease Ontology \autocite{vasilevsky_mondo_2022, shefchek_monarch_2020}; the Orphanet database on rare diseases and orphan drugs \autocite{weinreich_orphanet_2008}; the human protein-protein interaction (PPI) network, which includes PPI information from \textcite{menche_uncovering_2015}, BioGRID \autocite{oughtred_biogrid_2021}, STRING \autocite{szklarczyk_string_2023}, and the human reference interactome generated by \textcite{luck_reference_2020}; the Reactome pathway knowledge base \autocite{milacic_reactome_2024}; the Side Effect Resource (SIDER) for data on adverse drug reactions \autocite{kuhn_sider_2016, kuhn_side_2010}; the Uberon anatomical ontology \autocite{haendel_unification_2014, mungall_uberon_2012}; and the Unified Medical Language System (UMLS) knowledge base \autocite{bodenreider_unified_2004}.

These datasets were carefully analyzed in a semi-automated fashion to extract and retain only high-quality associations. Analysis details follow those described in \textcite{chandak_building_2023}, updated to the latest versions of the datasets at the time of analysis and automated for reproducibility. Briefly, human gene expression data from Bgee was subset to retain Uberon-coded anatomical entities, gold quality calls with an adjusted $p$-value of $\leq 0.01$, and data with an expression rank less than 25,000, producing 5,198,776 raw anatomy-gene associations where gene expression was either present or absent in certain tissues or cell types. CTD exposure data was similarly processed to obtain 206,586 associations; DisGeNET data contained 84,038 associations between genes and diseases or phenotypes; and DrugBank pharmaceutical data (version 5.1.10) yielded 2,867,056 synergistic drug-drug interactions and 27,250 drug-protein associations, including carrier, enzyme, target, and transporter relationships. The Drug Central database (version released on May 10th, 2023) was processed to extract 41,971 relationships among drugs and diseases, including 27,671 contraindications, 11,775 indications, and 2,525 records of off-label use. The Entrez database was processed with the \texttt{GOATOOLS} package \autocite{klopfenstein_goatools_2018} to yield 291,835 associations of genes with GO pathways, including 138,367 associations with biological processes, 83,125 associations with cellular components, and 70,343 associations with molecular functions. The \texttt{GOATOOLS} package was also used to process the GO database itself, producing 68,651 ontological relationships among these pathways. From the HPO ontology, 256,768 positive associations and 554 negative associations between diseases and phenotypes were constructed. The MONDO disease ontology was used to construct 37,810 disease-disease associations, while 2,647 pathway-pathway and 44,865 pathway-gene associations were constructed from the Reactome database. Finally, SIDER produced 202,736 drug-phenotype associations, while Uberon produced 24,800 anatomical relationships. After pre-processing, data harmonization was performed to map database-level objects with shared identifiers to common nodes in the KG and then construct edges of various types from relationships among those nodes. Additionally, 88 nodes with duplicate names were identified and harmonized. Altogether, from these original 20 databases, 143,186 nodes connected by 6,817,223 edges were constructed, forming the backbone of \kg.

In addition to these original 20 data resources, we also included 16 databases from the Open Targets Platform \autocite{ochoa_next-generation_2023, koscielny_open_2017}, a large-scale public-private partnership for drug target identification and prioritization that includes data about targets, diseases, phenotypes, and drugs. Datasets analyzed from Open Targets included the Cancer Gene Census \autocite{sondka_cosmic_2018}, ChEMBL (for both drug-disease and drug-gene edges) \autocite{mendez_chembl_2019, gaulton_chembl_2012}, Clinical Genome Resource (ClinGen) \autocite{strande_evaluating_2017}, Project Score \autocite{pacini_comprehensive_2024, behan_prioritization_2019}, CRISPR screens \autocite{tian_genome-wide_2021}, Expression Atlas \autocite{papatheodorou_expression_2020, papatheodorou_expression_2018}, Gene Burden, Gene2Phenotype (G2P) \autocite{thormann_flexible_2019}, Genomics England PanelApp \autocite{martin_panelapp_2019}, IntOGen \autocite{martinez-jimenez_compendium_2020}, Orphanet \autocite{weinreich_orphanet_2008}, PROGENy \autocite{schubert_perturbation-response_2018}, Reactome \autocite{milacic_reactome_2024}, SLAPenrich \autocite{iorio_pathway-based_2018}, systems biology gene signatures \autocite{mostafavi_molecular_2018, peters_functional_2017, huan_systems_2013, zhang_integrated_2013}, and UniProt literature \autocite{the_uniprot_consortium_uniprot_2025, the_uniprot_consortium_uniprot_2021}. These 16 Open Targets databases were analyzed to extract edges to enrich \kg. The processing steps and decisions taken when analyzing each database were as follows. Of note, Open Targets databases were used to construct edges rather than nodes. Thus, prior to any downstream filtering or analysis, candidate edges were first subset to those incident on nodes already in the KG.

First, from the Cancer Gene Census, 76,289 gene-disease associations were extracted, of which 71,791 were incident on nodes in the KG. These associations were scored according to the Cancer Gene Census tier system \autocite{sondka_cosmic_2018}, where the score of each gene-disease association is incremented or decremented by multiples of $0.25$ based on the strength of the supporting evidence. To extract only the associations of the highest strength and quality, we filtered for edges with scores $> 0.5$, indicating that the gene is mutated more frequently in that particular disease compared to other diseases, and that mutations in the gene occur more frequently than in other genes of similar length in the same disease. Next, 625,734 associations between drugs and diseases or genes were retrieved from the ChEMBL database of drug-like molecules -- including both compounds approved for marketing by the U.S. Food and Drug Administration (FDA) and clinical candidates -- of which 410,317 edges were incident on nodes in the KG. We then consulted Open Targets scores for each association, which were determined using a machine learning analysis that categorized each corresponding clinical trial into 17 possible reasons for termination. First, scores were assigned based on clinical precedence, \ie, scores of 0.05 were assigned for trials concluding before Phase I, 0.1 for post-Phase I termination, 0.2 for Phase II, 0.7 for Phase III, and 1 for Phase IV. Next, for trials that stopped early, scores were decremented by 0.5 if trials terminated due to negative outcomes, safety concerns, or side effects. We filtered for edges with scores $> 0.05$, then constructed edges of type ``weak clinical evidence'' for score $\in (0.05, 0.5)$, ``strong clinical evidence'' for score $\in [0.5, 1)$, and ``indication'' for score $ = 1$, ultimately producing 403,866 new edges in \kg.

The next dataset to be processed was the ClinGen Gene-Disease Validity Curation, from which 2,183 gene-disease associations were retrieved that described whether variance in a given gene is causative of a given disease. Of these, 2,039 were incident on edges in the KG. Edges were assigned scores based on the supporting evidence, where 0.01 indicates limited, refuted, disputed, or non-existent evidence; 0.5 suggests moderate evidence; and 1 indicates strong or definite evidence \autocite{strande_evaluating_2017}. Again, to retain only high-quality edges, we filtered for edges with scores of 1 for only strong or definite genetic associations, leaving 1,299 edges to be included in \kg. The Project Score database from the Wellcome Sanger Institute -- which analyzed 930 genome-wide CRISPR-Cas9 functional genomics screens to identify gene dependencies in 27 cancer types -- contained 1,838 gene-disease associations \autocite{pacini_comprehensive_2024}. Associations with Project Score priority scores greater than 41.5 were included as evidence, all of which were incident on edges in the KG. The CRISPR screens database contained 24,907 gene-disease associations, of which 21,721 were incident on gene and disease nodes in the KG. Associations were derived from genome-wide CRISPR interference, activation, and knock-out screens in differentiated human brain cell types, including glutamatergic neurons, astrocytes, microglia, and hematopoietic stem and progenitor cells. Following \textcite{tian_genome-wide_2021}, the 21,721 candidate edges were scored as the product of the phenotype score and $-\log_{10}(p\text{-value})$, representing both effect size and statistical significance. To extract only the most robust associations, we filtered for the 7,088 gene-disease edges with a score of 1, which included 1,830 upregulated (\ie, fold-change of the given gene's expression in the given disease is $> 0$) and 5,258 downregulated (\ie, fold-change $< 0$) edges. The Expression Atlas database contained 230,893 gene-disease associations, of which 197,107 were incident on gene and disease nodes in the KG. Duplicate edges were collapsed by computing the mean $\log_2(\text{fold-change})$ across multiple observations of the same gene-disease pair. To extract strong associations, we filtered for the 132,219 edges with fold-change of 2 (\ie, $|\log_2(\text{fold-change})| = 1$), which included 66,706 upregulated and 65,513 downregulated edges.

Next, we considered the Gene Burden dataset, which contains gene-disease edges derived from association tests in rare variant collapsing analyses, including and especially those related to neurological disease. Burden tests were integrated across various studies, including a whole-exome sequencing (WES) study of 454,787 UK Biobank (UKB) participants \autocite{backman_exome_2021}; the AstraZeneca PheWAS Portal, a WES study of 269,171 UKB participants \autocite{wang_rare_2021}, the Gene-Based Association Summary Statistics (Genebass) WES study of 394,841 UKB participants \autocite{karczewski_systematic_2022}; a whole-exome and whole-genome sequencing study of 42,607 autism spectrum disorder (ASD) cases in the Simons Foundation Powering Autism Research for Knowledge (SPARK) cohort \autocite{zhou_integrating_2022}; the Schizophrenia Exome Sequencing Meta-Analysis (SCHEMA) WES study of 24,248 individuals with schizophrenia and 97,322 controls from seven continental populations \autocite{singh_rare_2022}; the Epi25 Collaborative WES study of 9,170 individuals with epilepsy and 8,436 controls \autocite{feng_ultra-rare_2019}; the Autism Sequencing Consortium WES study of 35,584 individuals, including 21,219 family-based samples (6,430 ASD cases, 2,179 unaffected siblings, and both parents) and 14,365 case-control samples (5,556 ASD cases, 8,809 controls) \autocite{satterstrom_large-scale_2020}; a whole-genome and whole-exome sequencing study of 7,184 PD cases, 6,701 proxy cases and 51,650 controls from AMP-PD \autocite{makarious_large-scale_2023}; and several other studies \autocite{bomba_whole-exome_2022, akbari_multiancestry_2022, riveros-mckay_influence_2020}. A total of 27,381 gene-disease and gene-phenotype associations were retrieved from the Gene Burden dataset, of which 7,540 had gene and either disease or phenotype representations in the KG (where phenotypes were identified by the ``HP'' prefix in their Open Targets identifier). Edges integrated into \kg included 1,032 gene-phenotype edges and 6,508 gene-disease edges across 371 diseases.

After Gene Burden, we analyzed the G2P database of gene-disease panels curated from the literature by clinical geneticists. A total of 3,011 gene-disease or gene-phenotype associations were retrieved from G2P, of which 2,195 were incident on genes and either diseases or phenotypes in the KG. These associations were assigned the following evidence scores: 0.01 for limited evidence, 0.5 for moderate evidence, and 1 for strong or definitive evidence (or for genotypes associated with both the relevant disease of that edge as well as another disease that represents an incidental finding) \autocite{thormann_flexible_2019}. To select only the edges supported by strong or definitive evidence, we filtered for edges with scores of 1, leaving 1,992 gene-disease or gene-phenotype edges. Next, 31,394 gene-disease associations were retrieved from Genomics England PanelApp, a knowledge base of gene-disease relationships crowdsourced from and curated by clinical and scientific experts. Of the original 31,394 associations, 22,724 were incident on nodes in the KG. Gene-disease associations were assigned three possible classifications in Genomics England PanelApp based on the level of supporting evidence: red, indicating low confidence; amber, indicating moderate confidence; and green, indicating strong or diagnostic-grade confidence \autocite{martin_panelapp_2019}. We filtered for green-level associations, which require ``evidence from three or more unrelated families or from two or three unrelated families where there is strong additional functional data,'' leaving 20,324 gene-disease edges to be added to the KG. Finally, from the IntOGen, Orphanet, PROGENy, Reactome, SLAPenrich, systems biology gene signatures, and UniProt literature databases, 4,359, 6,342, 378, 10,095, 72,440, 389, and 4,135 associations were retrieved, respectively, of which 4,296, 5,156, all 378, 9,897, 72,272, 309, and 2,948 associations were incident on nodes in the KG and were thus included.

Of note, some Open Targets databases were excluded as they did not meet our inclusion criteria of \textbf{(1)} direct clinical or experimental evidence; \textbf{(2)} human data rather than animal models; \textbf{(3)} binary associations between two nodes in the KG; and \textbf{(4)} for genetic associations, gene-level rather than variant-level relationships (to prevent dominance of variant nodes in the KG). For example,  Cancer Biomarkers was excluded because it contains information about relationships between drugs, biomarkers, targets, and cancer subtypes, which would require the construction of hyperedges connecting several nodes in the KG rather than a binary edge connecting two nodes \autocite{tamborero_cancer_2018}. The EMBL-EBI Europe PubMed Central database was excluded as associations were derived from text co-occurrences in the literature identified by deep learning-based named entity recognition rather than experimental or clinical evidence \autocite{the_europe_pmc_consortium_europe_2015, kafkas_literature_2017}. The International Mouse Phenotypes Consortium data was excluded as associations were derived from phenotypic effects of gene knockouts in mice rather than humans \autocite{smedley_phenodigm_2013}. Finally, the ClinVar \autocite{shen_cmat_2024, cezard_european_2022, landrum_clinvar_2020, landrum_clinvar_2014} and UniProt variants \autocite{the_uniprot_consortium_uniprot_2025, the_uniprot_consortium_uniprot_2021} datasets were excluded because they contained variant-level rather than gene-level relationships. With the inclusion of the Open Targets Platform, the number of biomedical relationships in \kg increased to 7,048,887 edges.

In building \kg, the July 2023 versions of all databases were used, with the exception of Open Targets Platform, for which version 23.09 was used, retrieved in January 2024. Unless otherwise indicated, all code was written and executed in Python (version 3.11.10) and R (version 4.4.0).

\subsection{Contextualizing \kg to the human brain}\label{method:contexualize-kg-brain}

With the datasets described in Section~\ref{method:unifying-datasets} alone, \kg already included relevant information about the nervous system; for example, the Gene Burden dataset included whole-exome and whole-genome studies from individuals with ASD, schizophrenia, epilepsy, and PD among other conditions. To further capture the rich molecular interactions and unique cellular relationships relevant to neurological disease, we contextualized the KG to the brain by including data from approximately 3.7 million cells from single-cell RNA sequencing (scRNA-seq) atlases of the adult human brain.

First, we analyzed a single-nucleus RNA sequencing (snRNA-seq) atlas of the adult human brain \autocite{siletti_transcriptomic_2023}. This atlas was constructed by sequencing postmortem tissue from 606 samples in three adult human donors dissected from 112 anatomically-distinct locations across 14 brain regions -- including the telencephalon, diencephalon, midbrain, hindbrain, and cervical spinal cord. This dataset was also enriched for neurons using fluorescence-activated cell sorting. In \textcite{siletti_transcriptomic_2023}, Harmony \autocite{korsunsky_fast_2019} was used to integrate cells across donors; then, the Paris algorithm \autocite{bonald_hierarchical_2018} was used to build a dendrogram that defined 31 cell types (referred to by Siletti \etal as ``superclusters''), 461 subtypes, and 3,313 subclusters (Figure \ref{si:fig:siletti}a). Using the \texttt{scanpy} package \autocite{wolf_scanpy_2018} in Python, we analyzed this data from 2,480,956 neurons and 888,263 non-neuronal cells. Non-neuronal cell types included astrocytes, microglia, oligodendrocytes, oligodendrocyte precursor cells, ependymal cells, Bergmann cells, vascular cells, and choroidal epithelial cells. Briefly, mitochondrial genes were annotated, and quality control metrics were computed -- including the number of genes expressed in the count matrix, the total counts per cell, and the percentage of counts in mitochondrial genes. To remove cells with stress or damage, poor quality cells, and potential doublets or multiplets, 40,627 neuronal and 2,965 non-neuronal cells exceeding 12,000 total counts for neurons, 7,000 total counts for non-neuronal cells, or with greater than 5\% mitochondrial gene expression were removed. Next, each cell was normalized by total counts over all genes to 10,000 counts per cell, after which a logarithmic transformation was applied. All 29,921 human genes with valid Entrez gene identifiers were retrieved from BioMart \autocite{smedley_biomart_2009} and were used to filter the genes in the \textcite{siletti_transcriptomic_2023} snRNA-seq data. From the original 59,480 genes, 33,683 genes without Entrez gene identifiers, 1,158 genes with missing gene symbols, and 138 genes with duplicated gene symbols were removed. Finally, each gene was scaled to zero mean and unit variance, and values that exceeded a standard deviation of 10 were truncated. The final dataset consisted of 24,501 genes assayed across 2,387,555 neurons and 885,298 non-neuronal cells. The numbers of cells by each cell type are shown in Supplementary Table \ref{si:table:siletti_cell_counts}.

Next, differential expression analysis was performed to identify marker genes distinguishing 31 cell types, 461 subtypes, and 3313 subclusters. Marker genes were identified using a Student's $t$-test applied on the raw (\ie, unscaled) data \autocite{luecken_current_2019}. For cell types, marker genes were those with increased or decreased expression in each cell type relative to all other cell types; for subtypes, marker genes were those with differential expression in each subtype relative to all other subtypes from the same parent cell type; and for subclusters, marker genes were those with differential expression in each subcluster relative to all other subcluster with the same parent subtype. The numbers of subtypes and subclusters belonging to each supercluster are also shown in Supplementary Table \ref{si:table:siletti_cell_counts}.

We validated that cell type annotations in Siletti \etal were well-constructed by visualizing the expression of known marker genes of neurons (\textit{INA}), glutamatergic (\textit{SLC17A6}, \textit{SLC17A7}) and GABAergic (\textit{SLC31A1}) neurons, microglia (\textit{PTPRC}), astrocytes (\textit{AQP4}), oligodendrocytes (\textit{PLP1}), and oligodendrocyte precursor cells (\textit{PDFGRA}) (Supplementary Figures \ref{si:fig:siletti}b and \ref{si:fig:siletti}c). The expression of these literature-derived marker genes was visualized as scatter plots in the t-distributed stochastic neighbor embedding (t-SNE) basis, which were compared to Figure 1b of \textcite{siletti_transcriptomic_2023}. We also validated marker genes identified by the differential expression analysis using gold-standard markers of astrocytes and microglia previously published in the literature and in our prior work. For example, in Viejo and Noori \etal \autocite{viejo_systematic_2022}, we performed a systematic review of 306 astrocyte immunohistochemical studies in postmortem AD brains followed by bioinformatics analyses. Our review identified 196 proteins dysregulated in reactive astrocytes of AD brains, which we refer to as the ``ADRA'' protein set. Proteins in the ADRA protein set were also significantly upregulated or downregulated in astrocytes in Siletti \etal (Supplementary Figure \ref{si:fig:siletti}d); that is, astrocyte marker genes identified from the snRNA-seq human brain atlas were significantly enriched for known ADRA immunohistochemical markers of astrocytes as assessed by gene set enrichment analysis ($p = 1.612 \times 10^{-4}$, $\text{ES} = 0.545$, $\text{NES} = 1.623$) \autocite{subramanian_gene_2005, ritchie_limma_2015, korotkevich_fast_2021}.

Finally, the brain regions, brain structures, cell types, subtypes, and subclusters were used to define new nodes in \kg. Then, based on the differentially expressed marker genes and cell type proportions, new edges were created from brain regions to brain structures, cell types to cell subtypes, brain structures to cell subtypes, cell subtypes to cell subclusters, cell types to marker genes, cell subtypes to marker genes, and cell subclusters to marker genes (Supplementary Figure \ref{si:fig:siletti}a). First, new nodes were created for all 14 brain regions (\ie, the amygdaloid complex, basal forebrain, basal nuclei, cerebellum, cerebral cortex, claustrum, extended amygdala, hippocampus, hypothalamus, midbrain, myelencephalon, pons, spinal cord, and thalamus) and 112 brain structures (\ie, anatomically-distinct dissections in Siletti \etal), which were defined by the Allen Brain Atlas. Edges between each brain structure and their parent brain region were constructed, and human readable names of all 112 brain structure nodes were created by manual review using neuroanatomical reference atlases. Next, for each cell subtype, the proportions of cells of that subtype present across each brain structure were calculated and visualized in a histogram (Supplementary Figure \ref{si:fig:siletti}e). Based on this histogram, a membership threshold of 5\% was determined; therefore, edges were constructed between a cell subtype and a brain structure if at least 5\% of the total cells belonging to that subtype were present in that brain structure. A total of 1,820 edges were constructed between brain structures and cell subtypes. Additionally, 450 parent-child edges were constructed between cell types and cell subtypes, and 3,244 edges were constructed between cell subtypes and cell subclusters. Lastly, 12,334 edges were constructed between cell types and their top 400 marker genes, 89,288 edges were constructed between cell subtypes and their top 200 marker genes, and 160,634 edges were constructed between cell subclusters and their top 50 marker genes (Supplementary Figure \ref{si:fig:siletti}a). In sum, 3,851 new nodes and 267,882 edges were instantiated in \kg from this single-cell transcriptomic atlas of the adult human brain.

Although the snRNA-seq atlas published by Siletti \etal features rich information about the healthy human brain, we also sought to incorporate transcriptomic data from neurological disease settings into \kg. Therefore, we analyzed 202,810 nuclei from the postmortem substantia nigra pars compacta of ten donors with pathological midbrain DA neuron loss and a clinical diagnosis of either PD or Lewy body dementia (LBD), as well as 184,673 nuclei from eight age-matched and postmortem-interval-matched neurotypical donors \autocite{kamath_single-cell_2022}. Nuclei were enriched for those from dopaminergic neurons using fluorescence-activated nuclei sorting. As described in Kamath \textit{et al.}, differential expression analysis across eight cell types and 68 subtypes was performed using model-based analysis of single-cell transcriptomes (MAST), a two-part generalized linear model designed for scRNA-seq data \autocite{finak_mast_2015}. Fixed-effect covariates -- including log-transformed UMI counts, sex, mitochondrial read percentage per nucleus, and disease status -- were controlled for. Of note, for each cell type and subtype, this analysis identified differentially expressed genes (DEGs) both in comparison against all other cells (\ie, marker genes) and in comparison between PD or LBD brains and control brains (\ie, PD-related DEGs). Within each cell type and subtype, marker and PD-related genes were ranked by $z$-score, which integrates changes in the discrete and continuous components of expression.

The eight cell types included were astrocytes, microglia, endothelial cells, oligodendrocytes, oligodendrocyte precursor cells, excitatory neurons, inhibitory neurons, and dopaminergic neurons. As before, 68 edges were constructed between cell types and subtypes, 3,200 edges were constructed between cell types and their top 400 marker genes, and 13,600 edges were constructed between subtypes and their top 200 marker genes. In addition, 6,400 edges were constructed between cell types and the top 800 (\ie, 400 upregulated and 400 downregulated) PD-related DEGs in each cell type, and 26,800 edges were constructed between subtypes and the top 400 (\ie, 200 upregulated and 200 downregulated) PD-related DEGs in each subtype. Edges derived from disease-associated DEGs were assigned types based on their direction of effect, labeled as either ``upregulated in PD'' or ``downregulated in PD.'' Altogether, 76 new nodes and 50,068 new edges were constructed from the 387,483 nuclei in this snRNA-seq study of the neurodegenerating PD brain.

Finally, to remove isolated pockets of nodes or prune any nodes without edges connecting them to the rest of the KG, the largest weakly connected component of the KG was extracted using the \texttt{igraph} software package for network analysis. In the resulting connected component, 99.9368\% of the nodes and 99.9988\% of the edges were retained; this connected component was adopted as the final \kg knowledge graph. As \kg is undirected, reverse edges were added to ensure that relationships in \kg were bidirectional.

\section{\model architecture and optimization}\label{methods:model-architecture-optimization}

\subsection{Model architecture}\label{methods:model-architecture}

To learn representations on \kg, graph neural networks (GNNs) -- an extension of neural networks to the graph domain -- were used \autocite{scarselli_graph_2009}. \model learns weight matrices that define a series of differentiable functions mapping node and edge features to link prediction probabilities. During training, nodes update their vector representations, or embeddings, in each layer by aggregating messages received from their neighbors (Figure \ref{fig:overview}c). In a standard GNN, for node $u \in \scriptV$ with one-hop neighborhood $\scriptN_u$ and node features in the $(\ell-1)$-th layer as $\mathbf{x}_u^{\ell-1}$, the $\ell$-th message passing layer in a GNN can compute the new embedding $\mathbf{x}_u^\ell$ as follows \cite{bronstein_geometric_2021}:
\begin{gather*}
    \mathbf{x}_u^\ell = \rho\left(\mathbf{x}_u^{\ell-1}, \bigoplus_{v \in \mathcal{N}_u}{\psi(\mathbf{x}_u^{\ell-1}, \mathbf{x}_v^{\ell-1})}\right),
\end{gather*}
where $\rho$ and $\psi$ are the differentiable \textit{update} and \textit{message} functions, respectively, while $\bigoplus$ is a permutation invariant aggregation operator. Then, to produce $p(u, r, v)$, the final node embeddings in an $L$-layer GNN can be combined through a decoder function, such as the dot product $\mathbf{x}_u^L \cdot \mathbf{x}_v^L$.

For \model, two important considerations were made in the design of the GNN encoder: attention and relation heterogeneity. First, instead of using standard averaging or maximum pooling over all neighbors in the $\bigoplus$ aggregation operator, attention mechanisms can allow a node to update its representation by only considering important neighbors. That is, rather than treat all neighbors equally during message passing, each node can compute a weighted average of its neighbors' representations, where the weights, or attention scores $\alpha$, are adaptively learned. In this manner, each node in the GNN can focus on its most relevant neighbors in the KG \autocite{velickovic_graph_2018}. This approach is inspired by human cognitive systems that, when presented with large data streams, focus only on subsets of information as needed. Attention mechanisms -- and the popular transformer architecture \autocite{vaswani_attention_2017} that they have enabled -- have underpinned advances in deep learning on language, vision, and graphs \autocite{niu_review_2021}; thus, we sought to select a transformer-based GNN design with attention mechanisms.

It is also critical to recognize that \kg is a heterogeneous graph composed of different node types and edge types, given by the sets $\scriptA$ and $\scriptR$. This diversity of possible relations represents an important source of biomedical knowledge encoded in the graph. For instance, when predicting drug-disease relations to match candidate therapeutics to the diseases they may treat, not only do drugs and diseases in \kg share edges with non-drug and non-disease neighbors, but drug-disease relations themselves are of multiple types in the knowledge graph. These edge types include contraindications, indications, and off-label use. High contraindication likelihood suggests that a drug would not represent an attractive drug repurposing target, while indication and off-label use predictions suggest the opposite; however, traditional GNN methods for homogenous graphs would fail to differentiate between these three relationships. Thus, when learning both node embeddings and attention weights to model \kg, it is necessary to leverage the heterogeneity of possible relations in the KG and preserve the complex structural information encoded therein. As different types of nodes and edges encode different information, they may require separate representation spaces.

To incorporate both attention mechanisms and type-specific representation spaces in \model, we selected a heterogeneous graph transformer (HGT) \autocite{hu_heterogeneous_2020} architecture with three layers and 578,083,992 learnable parameters (Figure \ref{fig:overview}c). A multi-head attention mechanism was employed to allow nodes to independently learn different schemes to assess importance in their neighborhood. As described in \textcite{hu_heterogeneous_2020}, for nodes $u, v \in \scriptV$ connected by edge $(u, r, v) \in \scriptE$ and with node types $\phi(u), \phi(v) \in \scriptA$, the attention score $\alpha'_i(u,r,v)$ of the $i$-th attention head is computed by the HGT as follows:
\begin{gather*}
    \alpha'_i(u,r,v) = \frac{Q_i(u) \cdot \mathbf{W}^\alpha_r \cdot K_i(v) \cdot \mu_{r}}{\sqrt{d}} \\
    Q_i(u) = \mathbf{W}^{Q,i}_{\phi(u)} \cdot \mathbf{x}^{\ell-1}_u \\
    K_i(v) = \mathbf{W}^{K,i}_{\phi(v)} \cdot \mathbf{x}^{\ell-1}_v
\end{gather*}
Here, $Q_i(u)$ is the query vector of node $u$ , obtained by a node-type-specific linear transformation to its previous layer embedding, and $K_i(v)$ is the key vector of node $v$, similarly transformed based on type. $\mathbf{W}^\alpha_r \in \mathbb{R}^{\frac{d}{h}\times\frac{d}{h}}$ is the edge-type-specific attention transformation matrix associated with relation $r$, $\mu_{r}$ is an edge-type-specific prior weight, $d$ is the size of the hidden dimension of layer $\ell$, and $h$ is the total number of attention heads (therefore, $\frac{d}{h}$ is the dimensionality per head). To compute the final attention vector $\alpha(u)$ for node $u$ across all one-hop neighbors $v \in \scriptN_u$, the attention heads of each neighbor are concatenated and then normalized into a probability distribution:
\begin{gather*}
    \alpha(u) = \underset{v \in \scriptN_u}{\text{softmax}}\left(\Big\|_{i = 1}^{h} \alpha'_i(u,r,v)\right) = \frac{\exp\left(\Big\|_{i = 1}^{h} \alpha'_i(u,r,v)\right)}{\sum_{v'\in\scriptN_u}{\exp\left(\Big\|_{i = 1}^{h} \alpha'_i(u,r,v')\right)}}
\end{gather*}
Note that $\|$ indicates vector concatenation. 

Concurrently with the attention calculation, heterogeneous multi-head messages are also calculated. Recall that $\psi$ is the differentiable \textit{message} function. Then, the $i$-th message head, $\psi'_{i}(u,r,v)$, is computed by:
\begin{gather*}
    \psi'_{i}(u,r,v) = \mathbf{W}_{\phi(v)}^{M,i} \cdot \mathbf{x}_v^{\ell - 1} \cdot \mathbf{W}^\psi_r \\
    \psi(u,r,v) = \Big\|_{i = 1}^{h} \psi'_{i}(u,r,v)
\end{gather*}
Here, $\mathbf{W}_{\phi(v)}^{M,i}$ is the node-type-specific transformation matrix that projects the previous layer embedding of node $v$ into the message vector and $\mathbf{W}^\psi_r \in \mathbb{R}^{\frac{d}{h}\times\frac{d}{h}}$ is the edge-type-specific message transformation matrix associated with relation $r$. Again, messages are concatenated across all message heads to arrive at the final message $\psi(u,r,v)$ per node pair. Finally, the normalized attention-weighted average of messages across all neighbors $v \in \scriptN_u$ of node $u$ is calculated to update the message of that node:
\begin{gather*}
    \mathbf{x}_u^\ell = \sigma\left(\mathbf{W}_{\phi(u)}^{\rho}\sum_{v \in \scriptN_u}{\alpha(u, r, v)\cdot\psi(u, r, v)}\right) + \mathbf{x}_u^{\ell-1}
\end{gather*}

To compute the new representation $\mathbf{x}_u^\ell$ of node $u$ at the $\ell$-th hidden layer of the GNN, the \textit{update} function $\rho$ is applied. $\rho$ is parameterized by a node-type-specific transformation matrix $\mathbf{W}_{\phi(u)}^{\rho}$, and also includes an attention-weighted average as the aggregation operator $\bigoplus$, a non-linear activation $\sigma$, and a residual connection \autocite{he_deep_2016}. \model employs the leaky rectified linear unit (ReLU) activation function:
\begin{gather*}
    \sigma = \begin{cases} 
        x & \text{ if } x \geq 0 \\
        a x & \text{ if } x < 0
    \end{cases}
\end{gather*}
where $a = 0.01$ is a positive constant that allows a small, nonzero gradient for negative values of $x$. Finally, to compute the likelihood $p(u,r,v)$, \model uses a bilinear decoder \autocite{yang_embedding_2015} with a trainable weight matrix $\mathbf{W}_r$ and leaky ReLU function $\sigma$:
\begin{gather*}
    p(u, r, v) = \sigma((\mathbf{x}_u^L)^\top\; \mathbf{W}_r \; \mathbf{x}_v^L),
\end{gather*}

Layer normalization was applied \autocite{ba_layer_2016} between HGT convolutional layers:
\begin{gather*}
    x' = \frac{x - \text{E}[x]}{\sqrt{\text{Var}[x] + \varepsilon}}\times \gamma +\beta
\end{gather*}
Layer normalization standardizes the activities of neurons by computing the mean and variance of all inputs to the neurons in a layer and also provides each neuron with its own adaptive gain ($\gamma$) and bias ($\beta$) parameters.

\subsection{Self-supervised pre-training}\label{methods:self-supervised-pre-training}

Pre-training was formulated as a binary link prediction task where, for a given edge type $r \in \scriptR$ and nodes $u, v \in \scriptV$, \model learns to predict the existence or non-existence of the relationship $(u,r,v)$ in \kg.

\xhdr{Negative edge construction} For each positive (\ie, ground truth) edge in the training data, we also sample $k$ negative edges that likely do not exist in the KG. Let the training dataset be $\scriptD = \{(u,r,v)_i, y_i\}_{i=1}^{N}$, with $u,v \in \mathcal{V}$, $N = (k+1)\times|\scriptE|$ (where $k$ is the ratio of negative to positive samples), and $y_i$ is the label defined as:
\begin{gather*}
    y_i = \begin{cases} 
        1 & \text{ if } (u,r,v^+)_i \in \scriptE \\
        0 & \text{ if } (u,r,v^-)_i \notin \scriptE
    \end{cases}
\end{gather*}
Here, $u$ is the source node, $v^+$ is a positive target node such that the edge $(u,r,v^+) \in \scriptE$, and $v^-$ is a negative target node such that the edge $(u,r,v^-) \notin \scriptE$. More accurately, it is unlikely that $v^-$ is connected to $u$ in \kg since the set of negative target nodes is generated by uniform random sampling of all nodes in the KG; however, false negatives are possible. For \model pre-training, we use $k = 1$; thus, for each positive edge, we sample a single negative edge.

\xhdr{Objective function} We seek for \model to learn to score positive edges as $1$ and negative edges as $0$. To that end, a binary cross-entropy loss function was used. The likelihood for a specific setting of our parameters $\mathbf{w}$ is:
\begin{gather*}
    p(\{y_i\}_{i=1}^{N}\; |\; \mathbf{w}) = \prod_{i=1}^{N} p(y_i = 1)^{y_i} \cdot (1 - p(y_i = 1))^{1 - y_i}
\end{gather*}
As described above, $p(y_i = 1 | (u_i, r_i, v_i)) = \sigma((\mathbf{x}_u^L)^\top\; \mathbf{W}_r \; \mathbf{x}_v^L)$. Note that the term inside the product evaluates to $p(y_i = 1)$ if $y_i = 1$ and $(1 - p(y_i = 1)) = p(y_i = 0)$ if $y_i = 0$. Maximizing this likelihood is equivalent to minimizing the negative log-likelihood or the binary cross-entropy loss:
\begin{gather*}
    \mathcal{L}(\{y_i\}_{i=1}^{N}\; |\; \mathbf{w}) = -\frac{1}{N} \sum_{i=1}^{N} \left[ y_i \ln \sigma((\mathbf{x}_{u,i}^L)^\top \mathbf{W}_{r,i} \mathbf{x}_{v,i}^L) + (1 - y_i) \ln (1 - \sigma((\mathbf{x}_{u,i}^L)^\top \mathbf{W}_{r,i} \mathbf{x}_{v,i}^L) \right]
\end{gather*}

To prioritize signals from rare edge types, the loss function was inversely weighted by edge type prevalence. Let $|r_i|$ be the number of edges of relation type $r_i$ in the KG. Then, our loss function is:
\begin{gather*}
    \mathcal{L}(\{y_i\}_{i=1}^{N}\; |\; \mathbf{w}) = -\frac{1}{\sum_{r\in\scriptR} |r|} \sum_{i=1}^{N} |r_i| \left[ y_i \ln \sigma((\mathbf{x}_{u,i}^L)^\top \mathbf{w}_{r,i} \mathbf{x}_{v,i}^L) + (1 - y_i) \ln (1 - \sigma((\mathbf{x}_{u,i}^L)^\top \mathbf{w}_{r,i} \mathbf{x}_{v,i}^L)) \right]
\end{gather*}

\xhdr{Neighborhood sampling} In message-passing neural networks with $L$ layers, the embedding of any node recursively depends on its $L$-hop neighborhood. For an $L$-layer GNN with hidden state size $d$ trained on a KG with $|\scriptV|$ nodes, representing all hidden states requires $\text{O}(L\times d \times |\scriptV|)$ GPU memory. As \kg contains $|\scriptV| = 147,020$ nodes, the full message-passing graph cannot be tractably stored in memory all at once.

Instead, mini-batch sampling \autocite{bertsekas_incremental_1996} was employed, where for each step of gradient descent, 512 edges were selected to comprise a batch. Then, messages were only passed among the $L$-hop neighborhoods of the head and tail nodes upon which the edges of the batch are incident (as well as any negative edges constructed as described above); this subgraph of \kg is referred to as the message-flowing graph (MFG). Since \model has three layers, $L = 3$. Recall that \kg is undirected; therefore, an edge of type $r$ between nodes $u, v \in \scriptV$ is represented as both a forward edge $(u, r, v)$ and a corresponding reverse edge $(v, r^{-1}, u)$. To prevent information leakage, for any edge in the batch, its reverse edge was excluded from the MFG. This prevents \model from receiving direct information about the relationship it is trying to predict, prompting it to learn from structural patterns in the graph rather than memorize direct connections.

During neighborhood construction, the size of the MFG grows exponentially with $L$. Therefore, even for small $L$ -- and especially in dense graphs, see Supplementary Table \ref{si:table:neurokg_nodes} -- the MFG may quickly expand to encompass the entire graph, which is known as the ``neighbor explosion'' phenomenon \autocite{zeng_graphsaint_2020, balin_layer-neighbor_2023}. To solve this challenge, various neighborhood sampling algorithms have been proposed to stochastically subsample the MFG \autocite{hamilton_inductive_2017, chen_stochastic_2018, liu_bandit_2020, zhang_biased_2021}. Motivated by this prior work, we designed a heterogeneous neighborhood sampling algorithm that, at each layer of the MFG, samples nodes from the frontier until a fixed budget is reached per node type. To upsample rare node types, nodes are sampled proportional to the square root of their relative prevalence in the MFG. An implementation of this heterogeneous neighborhood sampling algorithm was made publicly available as a contribution to the Deep Graph Library, a Python package for deep learning on graphs (see pull request \href{https://github.com/dmlc/dgl/pull/6668}{\#6668} in the \href{https://github.com/dmlc/dgl}{\texttt{dmlc/dgl}} repository on GitHub).

\xhdr{Model instantiation} Embeddings of all nodes $u \in \scriptV$ were initialized with random values $\mathbf{x}_u^0 \in \mathbb{R}^{d_0}$ drawn from the Gaussian distribution $\mathcal{N}(0, 1)$ and biases were zeroed. Following Glorot initialization \autocite{glorot_understanding_2010}, learnable relation and attention weights $\textbf{w}$ were sampled from the Xavier Uniform distribution. That is, $\textbf{w} \sim\text{Uniform}(-q, q)$ where:
\begin{gather*}
    q = \gamma \times \sqrt{\frac{6}{|\text{input}| + |\text{output}|}}
\end{gather*}
Here, $|\text{input}|$ is the number of input neurons in a given layer, $|\text{output}|$ is the number of output neurons, and $\gamma$ is a gain factor. Since the Leaky ReLU activation function was used, $\gamma$ was defined as:
\begin{gather*}
    \gamma = \sqrt{\frac{2}{1 + a^2}}
\end{gather*}
where $a = 0.01$, the same small positive constant used for the activation function. This gain factor is used to scale the weights appropriately so that activations maintain a stable variance as they propagate through the network, preventing vanishing or exploding gradients.

\xhdr{Model pre-training} \model was constructed and trained on \kg using the PyTorch (version 2.3.0) \autocite{paszke_pytorch_2019}, PyTorch Lightning (version 2.5.6), and Deep Graph Library (version 2.2.1) software packages \autocite{wang_deep_2020}. To measure pre-training performance, the \kg training data was randomly partitioned into training (80\%), validation (15\%), and test (5\%) sets. Importantly, to prevent information leakage, corresponding forward and reverse edges were assigned to the same data partition. During pre-training, message passing was only permitted over training edges; during validation, message passing was permitted over both training and validation edges; and during evaluation on the independent test set, message passing was performed over all edges. 

\xhdr{Hyperparameter optimization} Hyperparameters were selected via a Bayesian hyperparameter optimization search using the Weights \& Biases \autocite{biewald_experiment_2020} experiment tracking platform (Supplementary Figure \ref{si:fig:pretraining_sweep}). The search aimed to maximize the area under the receiver operating characteristic curve (AUROC) measured on the validation set. Hyperparameter values were sampled as follows: embedding size $d_0 \in \{512, 1024, 2048\}$; hidden dimension of the first HGT convolutional layer $d_1 \in \{256, 512, 1024\}$; output dimensionality $\in \{64, 128, 256\}$; number of attention heads $h \in \{2, 4\}$; dropout probability $\sim \text{Uniform}(0.2, 0.5)$; learning rate $\varepsilon \sim \text{LogUniform}(1 \times 10^{-2}, 1 \times 10^{-5})$; and weight decay $\sim \text{LogUniform}(1 \times 10^{-3}, 1 \times 10^{-6})$ (Supplementary Figure \ref{si:fig:pretraining_sweep}d). Note that when sampling from a log-uniform distribution, the natural logarithm of the sampled values follows a uniform distribution within the specified range. For all trials, the Adam optimization algorithm was used \autocite{kingma_adam_2017}. During the first 10 epochs, the learning rate was linearly decayed with each epoch by a factor of $\frac{1}{3}$. Unless otherwise stated, a random seed of 42 was used for training and evaluation across all trials, including the final models.

Model performance was evaluated at both aggregate and edge-type-specific levels across the training, validation, and test sets. Accuracy, AUROC, average precision, F\textsubscript{1} score, and loss were measured. A total of 55 hyperparameter tuning trials were completed; for each trial, a model was trained for up to the earlier of 48 hours or three epochs. Each trial was assigned a unique alphanumeric ID. Performance on the validation set was assessed every quarter training epoch. An early stopping criteria was implemented based on maximum validation AUROC with a patience of four validation checks. Therefore, for any given trial, if validation performance did not improve over a full pass over \kg (including all positive edges and all negatively sampled edges), then pre-training was halted for that trial. The parameter counts of trained models ranged from less than 100 million to over 2 billion parameters.

The importance of each parameter with respect to the optimization objective was evaluated by training a random forest model to predict the validation AUROC from hyperparameter configurations across all 55 trials (Supplementary Figure~\ref{si:fig:pretraining_sweep}e). The feature importance of the trained random forest model was considered as a measure of hyperparameter importance. In addition, the linear correlation between each hyperparameter and validation AUROC was also calculated (Supplementary Figure~\ref{si:fig:pretraining_sweep}e). After the completion of all trials, the best-performing hyperparameter configuration was selected based on link prediction AUROC measured on the independent test set. This configuration was then used to train the final version of \model on the entire \kg graph.

As additional trials were instantiated, the optimization strategy improved model performance, as assessed by both validation accuracy and AUROC. For the best-performing trials, the training and validation AUROC and accuracy consistently increased over training, while the loss decreased (Supplementary Figures~\ref{si:fig:pretraining_sweep}b and \ref{si:fig:pretraining_sweep}c). Performance metrics are reported in Supplementary Table \ref{si:table:sweep_metrics}. Most trials achieved strong performance, with AUROC on the independent test set exceeding 0.90 and accuracy exceeding 82\%. The sampled hyperparameter configurations across all 55 trials are shown in Supplementary Figure~\ref{si:fig:pretraining_sweep}d, while the importance of each hyperparameter with respect to test AUROC, as well as correlation with test AUROC, is shown in Supplementary Figure~\ref{si:fig:pretraining_sweep}e.

The final model architecture was selected based on performance on the independent test set. We sought to identify a hyperparameter configuration that performed well across all edge types rather than only the most prevalent. Note that the average test AUROC can be biased by prevalent edge types in \kg. Therefore, model performance was evaluated at an edge-type-specific level; that is, for each edge type, all 55 trials were ranked based on test AUROC. Then, the top 3 trials per edge type with the highest performance were identified. Trials were ranked based on the number of edge types for which they achieve within the top 3 test AUROC performance (Supplementary Figure~\ref{si:fig:pretraining_sweep}a). This metric, which is unbiased by edge type prevalence, closely correlates with average test AUROC, with a Spearman's rank correlation coefficient of $\rho = 0.9057$. The best performing trial was \texttt{hvkp2hjs}, with 578 million parameters and the following hyperparameter configuration: embedding size $d_0 = 1024$, hidden dimension $d_1 = 256$, output dimension $= 128$, number of attention heads $h = 4$, dropout probability $= 0.4547$, learning rate $\varepsilon = 2.00 \times 10^{-4}$, and weight decay $= 5.80 \times 10^{-5}$. On the independent test set, trial \texttt{hvkp2hjs} reached 0.9145 AUROC, 82.23\% accuracy, 0.9085 precision, and 0.8223 F\textsubscript{1} score (Supplementary Table \ref{si:table:sweep_metrics}). This hyperparameter configuration was adopted as the final \model model and re-trained on the full KG.

\xhdr{Embedding generation} Finally, we computed and analyzed the learned embedding space of \model (Supplementary Note~\ref{si:note:emb-biomedically-organized}). For each node $u \in \scriptV$, the final representation $\mathbf{x}_u^L$ produced by message passing over the neighborhood $\scriptN_u$ using the three-layer HGT architecture (without the bilinear decoder) was extracted and considered as the embedding for that node. Embeddings were generated via message passing over the entire graph using a deterministic $L$-hop sampler from ShaDow-GNN \autocite{zeng_decoupling_2022}. To mitigate the ``neighbor explosion'' phenomenon, nodes with degrees greater than 4,000 were excluded. Of the 147,020 nodes in the KG, only 289 nodes (0.1966\%) were excluded at inference time, including 1 gene/protein node, 7 disease nodes, 1 molecular function node, 4 cellular component nodes, and 276 anatomy nodes.

\xhdr{Hardware} Models were predominantly trained on NVIDIA H100 HBM3 chips with 100 gigabytes (GB) CPU RAM and 80 GB GPU RAM within the AI Cluster at the Kempner Institute for the Study of Natural and Artificial Intelligence at Harvard University. In addition, NVIDIA A100 SXM4 chips with 40 GB GPU RAM within the Kempner AI Cluster were used, as well as NVIDIA Quadro RTX 8000 chips with 48 GB GPU RAM within the O2 high-performance computing (HPC) environment at Harvard Medical School. CPU-intensive computation -- for example, for inference or embedding generation -- was also performed on the O2 HPC cluster, including on high memory nodes with up to 750 GB RAM to support message passing over the KG.

\section{Experimental studies for \model evaluation}\label{methods:experimental-studies}

Next, we describe the experimental studies conducted to generate independent datasets for model evaluation. Note, \model makes predictions on nodes in \kg; therefore, it was necessary to map biomedical entities in the evaluation datasets to corresponding \kg nodes.

\subsection{
  \texorpdfstring{$\bm{\alpha}$-synuclein-associated screens}{Alpha-synuclein-associated screens}
}
\label{methods:asyn-studies}

We considered six $\alpha$-synuclein-associated experimental screens, including $\alpha$-synuclein interactome profiling screens and $\alpha$-synuclein genetic screens in humans or disease models. Top hits from these screens were reviewed and curated by PD experts.

\xhdr{MYTH system interactome} 260 proteins from a split-ubiquitin protein complementation assay using a membrane-based yeast two-hybrid (MYTH) system to detect $\alpha$-synuclein interactions at the membrane \autocite{chung_situ_2017, lam_rapid_2024}. 249 proteins (95.76\%) were represented in \kg.

\xhdr{APEX2 proximity labeling} 212 proteins from an $\alpha$-synuclein proximity-dependent biotinylation labeling assay using an engineered ascorbate peroxidase (APEX2) to capture proteins that are spatially co-localized with $\alpha$-synuclein in PD primary neurons \autocite{chung_situ_2017}. Proximity-dependent biotinylation has an effective radius of approximately 10 nm around the $\alpha$-synuclein bait protein \autocite{kim_probing_2014}; therefore, this method identifies direct binding interactors of $\alpha$-synuclein as well as indirect interactors in the intracellular neighborhood \autocite{go_proximity-dependent_2021}. 200 proteins (94.34\%) were represented in \kg.

\xhdr{SILAC immunoprecipitation mass spectrometry} 50 $\alpha$-synuclein interactors were identified using stable isotope labeling by amino acids in cell culture (SILAC) \autocite{hallacli_parkinsons_2022}. HEK 293 cells stably expressing GFP or $\alpha$-synuclein-GFP were metabolically labeled, lysed, and subjected to co-immunoprecipitation using anti-GFP antibodies. Immunoprecipitated proteins were digested with trypsin and analyzed via liquid chromatography-mass spectrometry. 45 proteins (90\%) were represented in \kg.

\xhdr{Targeted deep exome sequencing} 27 gene hits from a targeted exome screen to identify genetic variants that contribute to synucleinopathy risk~\autocite{nazeen_deep_2024}. Human homologs of 430 genetic overexpression modifiers of $\alpha$-synuclein and $\beta$-amyloid cytotoxicity in yeast (see below) as well as known PD, AD, and ataxia risk genes were sequenced in $n=496$ patients with synucleinopathies (\ie, PD, Lewy body dementia, and multiple system atrophy) at high depth (100x). 25 proteins (92.59\%) were represented in \kg.

\xhdr{Proteotoxicity genetic modifiers (yeast::human transposition)} A network of 437 genes linked to $\alpha$-synuclein proteotoxicity. 318 modifiers of $\alpha$-synuclein toxicity were identified through genome-wide screens of approximately 10,000 genetic interactions in yeast, including a deletion screen for nonessential genes that increase sensitivity to low levels of $\alpha$-synuclein and a galactose-inducible overexpression screen in yeast HiTox strains with over six copies of $\alpha$-synuclein~\autocite{khurana_genome-scale_2017}. These 318 modifiers were combined with an additional 14 modifiers derived from separate experiments in yeast and PD patient-derived neurons \autocite{gitler_-synuclein_2009, chung_identification_2013, caraveo_calcineurin_2014}. The total 332 genetic modifiers of $\alpha$-synuclein toxicity were mapped to human counterparts using the TransposeNet algorithm that considers sequence homology, structural similarity, and overlap in the interactome. Finally, in addition to the experimentally validated interactors, candidate modifiers predicted by the prize-collecting Steiner forest algorithm were also included. 413 genes (94.51\%) were represented in \kg.

\xhdr{Modifier gene overexpression (yeast::human transposition)} Hits overexpressed in genome-wide array screens in yeast to identify enhancement or suppression of $\alpha$-synuclein neurotoxicity \autocite{khurana_genome-scale_2017}. Modifiers of $\alpha$-synuclein toxicity (see above) were humanized using TransposeNet and expanded with the prize-collecting Steiner forest algorithm to yield 245 hits, all of which were represented in \kg.

\xhdr{Random genes} For reference, we also included a set of 500 genes randomly sampled from the 35,198 genes in \kg, excluding those already connected to PD or $\alpha$-synuclein in \kg.

\subsection{Essentiality screen in patient-derived dopaminergic neurons}\label{methods:essentiality-screen}

A guide RNA (gRNA) library of 19,993 genes was transduced into H9 human embryonic stem cells, which were then differentiated into midbrain dopaminergic neurons. During the differentiation process, expression of the Cas9 protein was induced via doxycycline addition, thereby knocking out constituent genes of the gRNA library in the pre-dopaminergic neural progenitors (Figure \ref{fig:pd}e). Whole-genome sequencing was performed on samples from differentiated neurons at days 26 and 42. gRNA representation was used to identify genes essential for neuronal survival with FDR-adjusted $p$-value $< 0.05$ and $\beta < -0.58$. After removing broadly essential genes not specific to dopaminergic neurons \autocite{wang_integrative_2019}, 693 dopaminergic essentiality genes remained; of these, 681 were represented in \kg.

\subsection{Pesticide-wide association study in PD}\label{methods:pd-pwas}

To evaluate \model's ability to predict neurotoxicity, we used data from a pesticide-wide association study (PWAS) \autocite{paul_pesticide_2023} conducted on PD patients ($n=829$) and controls ($n=824$) in the Parkinson's Environment and Genes study \autocite{ritz_pesticides_2016}. This work identified 68 pesticides associated with PD, which we refer to as ``PWAS hits.'' This study also tested the toxicity of ``PWAS hits'' in iPSC-derived midbrain dopaminergic neurons from a PD patient overexpressing wild-type $\alpha$-synuclein, identifying ten pesticides that led to cell death $> 3$ standard deviations above the DMSO control mean at a concentration of 30 $\mu$M. We mapped all 68 pesticides back to one of the 860 pesticide or exposure nodes in \kg. To do so, we applied a fuzzy string matching algorithm based on Levenshtein text edit distance on pesticide names, with a score cutoff of 80. The 68 PWAS hits were mapped to 41 matches in \kg composed of 33 unique nodes, and the 10 PWAS hits toxic to dopaminergic neurons were mapped to 6 matches in \kg with 5 unique nodes. One pesticide, trifluralin, was excluded to minimize data leakage as it was already connected to PD in the \kg pre-training data, leaving 32 pesticides overall and 4 pesticides toxic to dopaminergic neurons.

\subsection{Proteomic profiling of drug-treated human brain organoids}\label{methods:organoids}

\subsubsection{iPSC culture}

Generation and validation of the BD and control induced pluripotent stem cell (iPSC) lines was previously published \autocite{meyer_impaired_2024} (Supplementary Table~\ref{si:table:ipsc-lines-organoid}). The iPSC lines underwent stringent quality control to confirm pluripotency and a normal karyotype. This included alkaline phosphatase assay, karyotype analysis, and differentiation into the three germ layers. They were cultured feeder-free on Geltrex LDEV-Free Reduced Growth Factor Basement Membrane Matrix (Gibco) coated plates in mTeSR\textsuperscript{TM} Plus Basal Media (Stem Cell Technologies). Extended culturing and maintenance protocols are available at \url{http://doi.org/10.17504/protocols.io.261ge8b9jg47/v2}.

\subsubsection{Generation of cerebral organoids and drug treatment}\label{methods:generation-organoids-treatment}

Cerebral organoids were differentiated using an approach adapted from a previously described protocol with modifications \autocite{lancaster_cerebral_2013}. Briefly, 9,000 hiPSCs were seeded into each well of a Corning\textsuperscript{\textregistered} 96-well clear round-bottom ultra-low attachment microplate (Corning \#7007) to form embryoid bodies (EBs) in the presence of \SI{50}{\micro\Molar} ROCK inhibitor (Sigma \#Y-27632). When EBs reached \SI{600}{\micro\meter} in diameter, media was replaced with neural induction media for four days before embedding the EBs in Matrigel (Corning \#354234). Embedded EBs were grown in ultra-low attachment plates in differentiation media without vitamin A for five days without shaking. Differentiation media containing vitamin A was used for long-term maintenance with shaking on an orbital shaker at \SI{90}{\rpm}. Media was changed every 2–3 days.

Since \textit{in vitro} cultured cells lack the enzymes needed to metabolize precursor forms of vitamin D, we selected the physiologically active hormone calcitriol for treatment of both BD and control cortical organoids. Cortical organoids derived from $n = 5$ patient and $n = 4$ control iPSC lines were treated with \SI{5}{\nano\Molar} calcitriol, a concentration significantly lower than those reported in other \textit{in vitro} protocols \autocite{orme_calcitriol_2013, aljohri_neuroprotective_2019} to ensure sufficient biological activity during prolonged treatment. However, the concentration is higher than physiological calcitriol levels in CSF, which range from approximately \SI{0.005}{\nano\Molar} to \SI{0.094}{\nano\Molar} \autocite{balabanova_25-hydroxyvitamin_1984}. Organoids were treated with calcitriol starting on day 46. The drug was replenished every two days. After one week of treatment, three organoids per each iPSC line were pooled, washed, flash-frozen in liquid nitrogen, and processed for deep proteomic profiling. The extended protocol is available at \url{http://doi.org/10.17504/protocols.io.kxygx4mb4l8j/v3}.

\subsubsection{Tissue preparation and mass spectrometry analysis}

Organoid brain tissue was stored at \SI{-80}{\celsius} in \SI{2}{\milli\liter} Eppendorf tubes.

\xhdr{Protein extraction protocol} Samples were kept on dry ice, and the frozen brain tissue was ground in the tube using a pestle. \SI{130}{\micro\liter} of \SI{5}{\percent} SDS was added to the ground tissue. Samples were centrifuged at \SI{15,000}{g} for \SI{5}{\minute} using an Eppendorf Centrifuge 5420. The supernatant (\SI{100}{\micro\liter}) was transferred to new \SI{2}{\milli\liter} Eppendorf tubes, while the remaining \SI{30}{\micro\liter} was stored at \SI{-80}{\celsius} for future work.

\xhdr{Reduction, alkylation, and acidification} \SI{8.7}{\micro\liter} of \SI{10}{\milli\Molar} TCEP was added to each sample. Samples were then shaken and mixed for \SI{30}{\minute} at \SI{350}{rpm} on an Eppendorf ThermoMixer C. Subsequently, \SI{8.7}{\micro\liter} of \SI{10}{\milli\Molar} IAA was added to the reduced samples, which were shaken and mixed for an additional \SI{10}{\minute} at \SI{350}{rpm}. Then, \SI{21.7}{\micro\liter} of \SI{12.5}{\percent} phosphoric acid was added.

\xhdr{Trapping and cleaning proteins} \SI{1522}{\micro\liter} of binding and wash buffer (\SI{100}{\milli\Molar} TEAB in \SI{90}{\percent} methanol) was added to the sample and immediately vortexed; the sample visually appeared as a milky colloidal suspension. The suspension was transferred to an ProtiFi S-Trap\textsuperscript{TM} mini spin column and centrifuged at \SI{4000}{g} for \SI{30}{\second} (Eppendorf Centrifuge 5420), and the flow-through was discarded. Proteins were washed on the S-Trap mini spin column with \SI{1522}{\micro\liter} binding and wash buffer, followed by centrifugation at \SI{4000}{g} for \SI{30}{\second}. This wash step was repeated three times, and each flow-through was discarded. Triple washing at this step is critical to ensure SDS removal from the protein sample.

\xhdr{Digestion} A stock solution of Trypsin Platinum, Mass Spectrometry Grade (Promega) was prepared at \SI{100}{\micro\gram\per\milli\liter} in \SI{50}{\milli\Molar} TEAB. Stock trypsin was added to each S-Trap Mini column at a 1:50 (trypsin:protein) ratio. Samples were incubated for \SI{2}{\hour} at \SI{50}{\celsius} with shaking and mixing at \SI{350}{rpm} on an Eppendorf ThermoMixer C.

\xhdr{Peptide elution} \textit{Elution 1:} After digestion was completed, \SI{80}{\micro\liter} of \SI{50}{\milli\Molar} TEAB in water (pH 8.5) was added to each S-Trap mini spin column and centrifuged at \SI{4000}{g} for \SI{1}{\minute}; the peptide-containing flow-through was collected. \textit{Elution 2:} \SI{80}{\micro\liter} of \SI{0.2}{\percent} formic acid was added directly to each spin column and centrifuged at \SI{4000}{g} for \SI{1}{\minute}; the flow-through was collected. \textit{Elution 3:} \SI{80}{\micro\liter} of \SI{50}{\percent} acetonitrile in water was added directly to each spin column and centrifuged at \SI{4000}{g} for \SI{1}{\minute}; the flow-through was collected. For each sample, the three eluates were combined. All samples were dried using an Eppendorf Vacufuge Plus and resuspended in \SI{6}{\micro\liter} of \SI{0.1}{\percent} formic acid in ultrapure HPLC-grade water before LC-MS/MS injection.

\xhdr{Mass spectrometry}  
After separation, each fraction was subjected to a single LC-MS/MS experiment on an Orbitrap Astral Mass Spectrometer (Thermo Scientific) equipped with a NEO (Thermo Scientific) nanoHPLC pump. Peptides were separated on a \SI{300}{\micro\meter} × \SI{5}{\milli\meter} PepMap C18 trapping column (Thermo Scientific, Lithuania) followed by a PepSep \SI{75}{\micro\meter} × \SI{150}{\milli\meter} analytical column (Bruker, USA). Separation was achieved by applying a gradient of 5–25\% acetonitrile in \SI{0.1}{\percent} formic acid over \SI{45}{\minute} at \SI{250}{\nano\liter\per\minute}.  

Electrospray ionization was enabled by applying \SI{1.8}{\kilo\volt} using a PepSep electrode junction at the end of the analytical column, spraying from a stainless steel PepSep emitter SS \SI{30}{\micro\meter} LJ (Odense, Denmark). The Exploris Orbitrap was operated in data-dependent mode. The survey scan was performed in the Orbitrap across an \SIrange{400}{900}{m/z} range at a resolution of 1.2 $\times$ 10\textsuperscript{5}, followed by selection of DIA \SI{2}{\dalton} windows in the TOF section of the Orbitrap for MS2 spectra collection. The fragment ion isolation width was \SI{2}{m/z}, AGC was \SI{250}{\percent}, maximum ion time was \SI{5}{\milli\second}, and normalized collision energy was \SI{25}{\volt}.

\xhdr{Mass spectrometry data analysis}  
Raw data were analyzed using Proteome Discoverer 3.2.023 (Thermo Scientific) with CHIMERYS\texttrademark and PEAKS Studio 12.5 (Toronto, Canada). Assignment of MS/MS spectra was performed by searching against a protein sequence database including all entries from the Human UniProt database (SwissProt, 19,768 entries, 2019) and other known contaminants, such as human keratins and common laboratory contaminants. Database searches used a \SI{10}{\ppm} precursor ion tolerance and required peptide N and C termini to adhere to trypsin protease specificity, allowing up to two missed cleavages.  Carbamidomethylation of cysteine amino acids (+57.021464 Da) was set as a static modification, and methionine oxidation (+15.99492 Da) as a variable modification.  A 1\% false discovery rate (FDR) at the protein level was achieved via a target-decoy database search. Filtering was performed using Percolator (64-bit version) \autocite{kall_non-parametric_2008}.

Quantitative proteomics data were pre-processed in Python using the Wyss Analysis Software for Proteomics (WASP) Version 3. Briefly, peptide-level abundances were filtered for intragroup completeness, normalized by Trimmed Mean of M-values (TMM), imputed by left-tail sampling, re-filtered by peptide intensity and intragroup CV, and aggregated to the protein level by the mean of the top three most intense peptides. The differential expression analysis was done in R using the \texttt{limma} package (version 3.64.1).

\section{\model evaluation protocols}\label{methods:evaluation-protocols}

The experimental results described in Section~\ref{methods:experimental-studies} serve as ground truth data for assessing \model's performance. In Section~\ref{methods:evaluation-protocols}, we detail the protocols developed to use these and other datasets for model evaluation.

\subsection{Fine-tuning \model for PD pesticide prediction}\label{methods:pesticide-ft}

A fine-tuning task was implemented to predict which pesticides described in Section~\ref{methods:pd-pwas} were PWAS hits. To avoid overfitting on the small number of labeled observations, we performed principal component analysis on \model embeddings for all pesticide nodes in \kg. The top six components were used as input features for a classifier consisting of a two-layer feed-forward neural network with a ReLU activation and a final sigmoid output. The model was trained to predict the binary label of PWAS association using a binary cross-entropy loss function. The network was fine-tuned for 400 epochs with a batch size of 8 using the Adam optimizer with a learning rate of $1 \times 10^{-3}$.

\subsection{Constructing disease-centric data splits}\label{methods:disease-splits}

To construct the disease-centric data splits, for each disease in the evaluation set, we removed all drug-disease edges involving that disease from \kg before training. However, immediate neighborhood edge deletion may not be sufficient to prevent data leakage. For example, in the case of PD, even if direct drug–PD edges are removed, \model could still access edges linked to closely related conditions such as postencephalitic Parkinsonism, juvenile-onset Parkinson's, or Parkinsonian-pyramidal syndrome, which may share treatment strategies. Therefore, we aimed to remove drug-related information not only for each of the 17 evaluation diseases but also for related diseases that could leak indirect information into the training process. However, defining disease similarity is inherently challenging as diseases can be related along many biological, clinical, and semantic dimensions. To address this, we designed a two-stage approach (Supplementary Figure~\ref{si:fig:disease_similar_preds}b). In the first stage, all 377,400 possible non-self disease-disease pairs (\ie, $D\times d - d$, where $D =$ 22,201 diseases in \kg and $d = 17$ evaluation diseases) were assessed using three lightweight nomination models to identify candidate related diseases. These models were: \textbf{(1)} cosine similarity of disease name embeddings generated using Clinical BioBERT \autocite{alsentzer_publicly_2019}; \textbf{(2)} Levenshtein token set ratio, a string similarity metric based on edit distance between disease names; and \textbf{(3)} one-hop neighborhood overlap in \kg, quantified using the Jaccard similarity index. Clinical BioBERT is a biomedical large language model \autocite{devlin_bert_2019} trained on PubMed abstracts (4.5B words), PubMed Central full-text articles (13.5B words) \autocite{lee_biobert_2020}, and clinical text from 2 million notes (approximately 880 million words) in the MIMIC-III EHR database \autocite{johnson_mimic-iii_2016}. Disease pairs were nominated for further review if they exceeded empirically determined thresholds on any of the three metrics: Clinical BioBERT similarity $> 0.98$, Levenshtein token set ratio $> 80$, or Jaccard neighborhood similarity $> 0.1$.

In the second stage, nominated disease pairs were reviewed by GPT-4o \autocite{openai_gpt-4o_2024}, an LLM used as a semantic rater. GPT-4o was prompted (see Supplementary Figure~\ref{si:fig:disease_similar_preds}c) to score each candidate disease on a scale from 1 to 5 based on its clinical, biological, and therapeutic similarity to the held-out evaluation disease. Of the 615 nominated diseases across all 17 disease splits, 453 received an LLM similarity rating $\geq 3$ and were retained; the remaining 162 received scores $< 3$ and were excluded. For example, the BD split included: alcohol abuse, alcohol dependence, bipolar depression, bipolar disorder, bipolar I disorder, bipolar II disorder, cocaine dependence, depressive disorder, dysthymic disorder, endogenous depression, epilepsy, major affective disorder, major depressive disorder, manic bipolar affective disorder, schizophrenia, and unipolar depression (Figure~\ref{fig:bd}a).

Finally, although we initially constructed disease-centric splits for 25 neurological diseases, several of these diseases had insufficient drug evidence for meaningful evaluation. Specifically, we excluded any disease that had fewer than five drugs supported by indication, strong clinical evidence, or off-label use evidence. After applying this criterion, 17 neurological diseases remained and constituted the final evaluation set.

\subsection{\model prediction of a drug candidate for BD}\label{methods:bd-candidate-pred}

To identify a drug candidate for testing in human brain organoids, \model first performed an \textit{in silico} screen to rank all 8,160 drugs in \kg by the likelihood of indication for BD. The top 200 drugs ranked by \model for BD were then reviewed by GPT-4o \autocite{openai_gpt-4o_2024}, which scored each prediction from 1 (undesirable) to 5 (desirable) along four axes, with the following prompts:

\xhdr{Likelihood of efficacy} ``How likely is it that \texttt{\{drug\}} would be effective in treating bipolar disorder, based on current biomedical knowledge? Rate from 1 (very unlikely) to 5 (very likely).''

\xhdr{Novelty} ``How novel is the idea of using \texttt{\{drug\}} to treat bipolar disorder? Rate from 1 (already well-established) to 5 (completely novel, with no prior reports).''

\xhdr{Contraindication likelihood} ``Is \texttt{\{drug\}} contraindicated for patients with bipolar disorder? Rate from 1 (definitely contraindicated) to 5 (definitely not contraindicated).''

\xhdr{Mechanistic rationale} ``How strong is the mechanistic rationale for using \texttt{\{drug\}} to treat bipolar disorder? Rate from 1 (no known mechanism) to 5 (clear and well-supported mechanism).''

The LLM was instructed to answer with a single number from 1 to 5. Finally, \model rankings and LLM scores were reviewed by a panel of domain experts to identify notable candidates. Experts were provided with \model rankings for all 200 drugs, optionally filtered by LLM output based on empirically determined thresholds: likelihood score $\geq 2$, novelty score $\geq 3$, contraindication score $\geq 3$, and mechanistic rationale score $\geq 2$. Filtering by these thresholds yielded 57 drugs for human review, from which calcitriol was selected.